\documentclass{aastex}          
\usepackage{spr-astr-addons}    

\begin{document}
%
  \title{A study of the cross-correlation and time lag in black hole X-ray binary XTE J1859+226}

\shorttitle{Cross-correlation and time lag in XTE J1859+226}

\shortauthors{Pei et al.}

\author{Songpeng Pei\altaffilmark{1,2}}
\and
\author{Guoqiang Ding\altaffilmark{1}}
\and
\author{Zhibing Li\altaffilmark{1}}
\and
\author{Yajuan Lei\altaffilmark{3}}
\and
\author{Rai Yuen\altaffilmark{1,4}}
\and
\author{Jinlu Qu\altaffilmark{5}}

\email{peisongp@xao.ac.cn}

\altaffiltext{1}{Xinjiang Astronomical Observatory, Chinese Academy of Sciences, 150, Science
1-Street, Urumqi, Xinjiang 830011, China}

\altaffiltext{2}{University of Chinese Academy of Sciences, 19A Yuquan Road, Beijing 100049, China}

\altaffiltext{3}{National Astronomical Observatories, Chinese Academy of Sciences, Beijing 100012, China}

\altaffiltext{4}{Key Laboratory of Radio Astronomy, Chinese Academy of Sciences, Urumqi, Xinjiang, 830011, China}

\altaffiltext{5}{Key Laboratory for Particle Astrophysics, Institute of High Energy Physics, Chinese Academy of Sciences, Beijing 100049, China}

\begin{abstract}

With Rossi X-ray Timing Explorer (RXTE) data,we systematically study the cross-correlation and time lag in all spectral states of black hole X-ray binary (BHXB) XTE J1859+226 in detail during its entire 1999$-$2000 outburst that lasted for 166 days.
Anti-correlations and positive correlations and their respective soft and hard X-ray lags are only detected in the first 100 days of the outburst when the luminosity is high. This suggests that the cross-correlations may be related to high luminosity.
Positive correlations are detected in every state of XTE J1859+226, viz., hard state, hard-intermediate state (HIMS), soft-intermediate state (SIMS) and soft state.
However, anti-correlations are only detected in HIMS and SIMS, anti-correlated hard lags are only detected in SIMS, while anti-correlated soft lags are detected in both HIMS and SIMS.
Moreover, the ratio of the observations with anti-correlated soft lags to hard lags detected in XTE J1859+226 is significantly different from that in neutron star low-mass X-ray binaries (NS LMXBs).
So far, anti-correlations are never detected in the soft state of BHXBs but detected in every branch or state of NS LMXBs.
This may be due to the origin of soft seed photons in BHXBs is confined to the accretion disk and, for NS LMXBs, from both accretion disk and the surface of the NS.
We notice that the timescale of anti-correlated time lags detected in XTE J1859+226 is similar with that of other BHXBs and NS LMXBs.
We suggest that anti-correlated soft lag detected in BHXB may result from fluctuation in the accretion disk as well as NS LMXB.

\end{abstract}


\section{Introduction}
X-ray binaries (XRBs) are mass-exchange binary systems consist of a compact object (an accreting object primary) and a companion star (a nondegenerate secondary star). The X-rays are produced by the compact object (the accretor) accretes falling matter from the companion star (the donor). Based on the mass of the donor star, XRBs are further subdivided into low-mass X-ray binaries (LMXBs) ($M_{donor}\leqslant1\,\rm M_\odot$) and high-mass X-ray binaries (HMXBs) ($M_{donor}\geqslant10\,\rm M_\odot$) \citep{2006csxs.book.....L,2009ASSL..359.....C}. Depending on whether the compact object is a black hole (BH) or a neutron star (NS), XRBs can be further subdivided into black hole X-ray binaries (BHXBs) and NS XRBs. Based on the spectral properties and color-color diagrams (CCDs), NS LMXBs can be classified into two subtypes, viz., atoll sources and Z-sources \citep{1989A&A...225...79H}.

It is well known that the X-ray energy spectra of BHXBs often consist of a thermal (soft) and a non-thermal (hard) component. The thermal component, which originates from the inner accretion disk, is well modeled by a multitemperature blackbody. The non-thermal component is usually modeled as a power law (PL), and it is generally believed that it results from the Comptonization of soft photons (thermal component) in a hot corona (Compton cloud) and from the jet \citep{2004ApJ...609..972M, 2005ApJ...635.1203M}.

Based on the evolution of spectral and timing properties, the spectral states of BHXBs can be divided into five states: quiescent state, low/hard state (LHS), intermediate state (IMS), high/soft state (HSS), and very high state (VHS) \citep{2005AIPC..797..231R, 2006ARA&A..44...49R, 2010LNP...794...53B}.
\citet{2006csxs.book..157M}, however, used a quantitative three-state description to define the spectral states of BHXBs. The three states are hard, thermal, and steep power law (SPL) states, which are comparable with LHS, HSS and VHS, respectively \citep{2007A&ARv..15....1D, 2010LNP...794...53B, 2006ARA&A..44...49R}.
VHS and IMS are at different luminosities, while they are usually taken as one and the same transition state between the LHS and HSS. This is because they have similar spectral and timing behaviors, and their amalgamative state can be divided into hard-intermediate state (HIMS) and soft-intermediate state (SIMS) \citep{2007A&ARv..15....1D, 2010LNP...794...53B}.
Most states are over wide and overlapping in luminosity, which means that `low', `high' and `very high' have lost most of their significance, so the spectral states of BHXBs in the outburst can be divided into four basic states: hard state, HIMS, SIMS, and soft state \citep{2005Ap&SS.300..107H, 2008BASI...36..151D, 2009MNRAS.400.1603M, 2010MNRAS.403...61D, 2012A&A...542A..56N}.

Spectral and timing analyses are the most important means to study XRBs. In particular, the timing analysis has been widely used for investigation of the underlying geometrical structure and dynamics of the accretion disk and the radiation mechanism of XRBs.
Cross-correlation analysis is an important tool to carry out timing study from which cross-correlation functions (CCFs) and time lags between different energy bands are obtained.
Positive correlations (anti-correlations) between low and high energy band light curves correspond to positive (negative) cross-correlation coefficients (CCCs).
The correlation between two different energy band light curves is defined as ambiguous correlation when their CCFs do not present obvious correlation.
Soft time lag implies that the higher energy photons lead the lower energy photons, whereas hard time lag means that the higher energy photons reach behind.
If hard time lag is detected in one observation owning the anti-correlation, that hard time lag is called anti-correlated hard lag, and the reverse is anti-correlated soft lag. Similarly, the time lags can be divided into positive correlation hard lags and positive correlation soft lags for observations with positive correlation.

\citet{2004ApJ...616L.143C} first performed cross-correlation analysis on BHXB Cyg X-3 to study the correlations and time lags between the low and high X-ray energy bands.
They first detected anti-correlated long time-scale hard X-ray lags in the hard state of Cyg X-3, which range from hundreds of seconds to about one thousand seconds.
Since then cross-correlation analysis has been widely used for study of the timing properties of XRBs.
Tens of seconds to over 1500 seconds anti-correlated time lags were detected in the IMS of other BHXBs that are GRS 1915+105 \citep{2005ApJ...631.1072C}, XTE J1550-564, 4U 1630-47 \citep{2007ApJ...661.1055S}, H1743-322 \citep{2009RAA.....9..901S} and GX 339-4 \citep{2010ApJ...725.1317S}.
Similar kinds of anti-correlated time lags were observed in six other NS LMXBs, including four Z-sources (Cyg X-2 \citep{2008ApJ...677..461L}, GX 5-1 \citep{2012ApJS..200...16S}, XTE J1701-462 \citep{2014MNRAS.440.3726W} and GX 349+2 \citep{2016MNRAS.455.2959D}) and two atoll sources (4U 1735-44 \citep{2013AJ....146...60L} and 4U 1608-52 \citep{2014AJ....147...67L}), and positive correlated time lags were also observed in them.
Furthermore, the evolution of the cross-correlation along the branches in the CCD or hardness intensity diagram (HID) was studied for the six NS LMXBs.
Based on the systematic cross-correlation analysis of 4U 1608-52 (atoll source), \citet{2014AJ....147...67L} proposed that the distribution of the cross-correlations on the CCDs might be in evolution with the luminosity of the source. Similar results were also found in a Z-source XTE J1701-462 by \citet{2014MNRAS.440.3726W}.
However, the evolution of cross-correlations along with the luminosity of the source and the spectral states has never been investigated in BHXBs, nor has the cross-correlation analysis been performed on a whole outburst of BHXBs.

XTE J1859+226 is a BHXB which was first detected by the All-Sky Monitor (ASM) of Rossi X-Ray Timing Explorer (RXTE) during its outburst on Oct 9, 1999 (MJD 51460) \citep{1999IAUC.7274....1W}.
\citet{2009ApJ...699..453S} used correlation scaling method to upgrade the source distance from 6$-$11 kpc \citep{2002MNRAS.331..169H, 2002MNRAS.334..999Z} to 4.2 $\pm$ 0.5 kpc with the state transition data of RXTE observations.
Based on the optical photometry and spectroscopy observations of transient source XTE J1859+226, \citet{2011MNRAS.413L..15C} obtained an orbital period of 6.58 $\pm$ 0.05 hr, radial velocity amplitude of $K_2= 541 \pm 70~\rm km\,s^{-1}$ for the companion star, mass function $f(M)=4.5 \pm 0.6~\rm M_\odot$, and a lower mass limit of $5.42\,\rm M_\odot$.
\citet{1999IAUC.7274....2M} detected low-frequency quasi-periodic oscillations (LFQPOs) and \citet{2000ApJ...535L.123C} discovered high-frequency quasi-periodic oscillations (HFQPOs) in the X-ray intensity of XTE J1859+226.
Using the RXTE/BeppoSAX data, \citet{2013MNRAS.428.3295F} discovered strong positive correlation between the root mean square (RMS) and the fraction of Comptonized seed photons.
Using RXTE data obtained during the whole 1999$-$2000 outburst, \citet{2014AdSpR..54.1678R} studied the evolution of spectral and temporal properties associated with different spectral states to explore the accretion dynamics. Results from study for the variation of spectral and temporal properties show that the whole outburst of XTE J1859+226 evolved through seven state transitions: hard state $\rightarrow$ HIMS $\rightarrow$ SIMS $\rightarrow$ soft state $\rightarrow$ SIMS $\rightarrow$ HIMS $\rightarrow$ hard state \citep{2014AdSpR..54.1678R}.

The motivation for this work is to study in detail the evolution of cross-correlations and long-term time lags along with the spectral states of BHXBs. The evolution of correlations and long-term time lags along with the spectral states of BHXBs has never been investigated, and we select BHXB XTE J1859+226 as it evolved through a sequence of spectral states during the 1999$-$2000 outburst \citep{2014AdSpR..54.1678R}.
The observations and data reduction are described in Section 2. The results are shown in Section 3, and discussion is present in Section 4. Finally, our conclusions are drawn in Section 5.

\section[]{OBSERVATIONS AND DATA REDUCTION}

In HEASARC database, there are 129 available observations generated by the two onboard detectors of RXTE, viz., Proportional Counter Array (PCA) and High Energy X-ray Timing Experiment (HEXTE). We excluded observations with exposure time less than 1000 s and two observations that are out of the outburst. The remaining 114 observations span over 164 days from October 11, 1999 (MJD 51462.76) to March 23, 2000 (MJD 51626.59).
Based on the 114 observations, we study the characteristics of XTE J1859+226 covering the entire duration of the single outburst.
Given that the light curves from some observations are discontinued, we follow \citet{2012ApJS..200...16S, 2013AJ....146...60L, 2014AJ....147...67L, 2014MNRAS.440.3726W} by dividing these observations into several individual segments with uninterrupted duration ($>$ 1000 s). We finally obtain 140 observations/segments.

\begin{figure*}
\centering
\includegraphics[width=7.6cm]{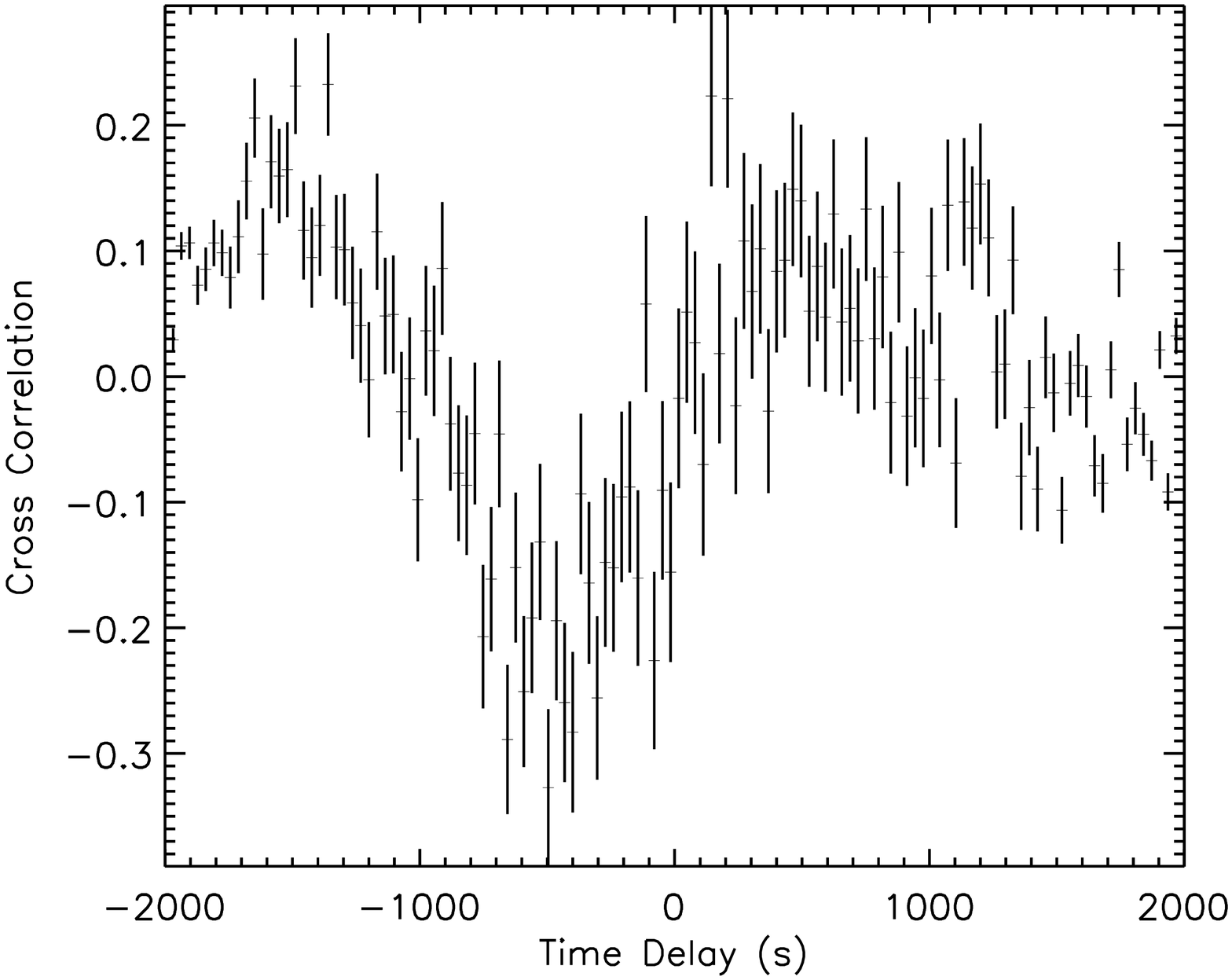}
\hspace{-0.62em}
\includegraphics[width=7.6cm]{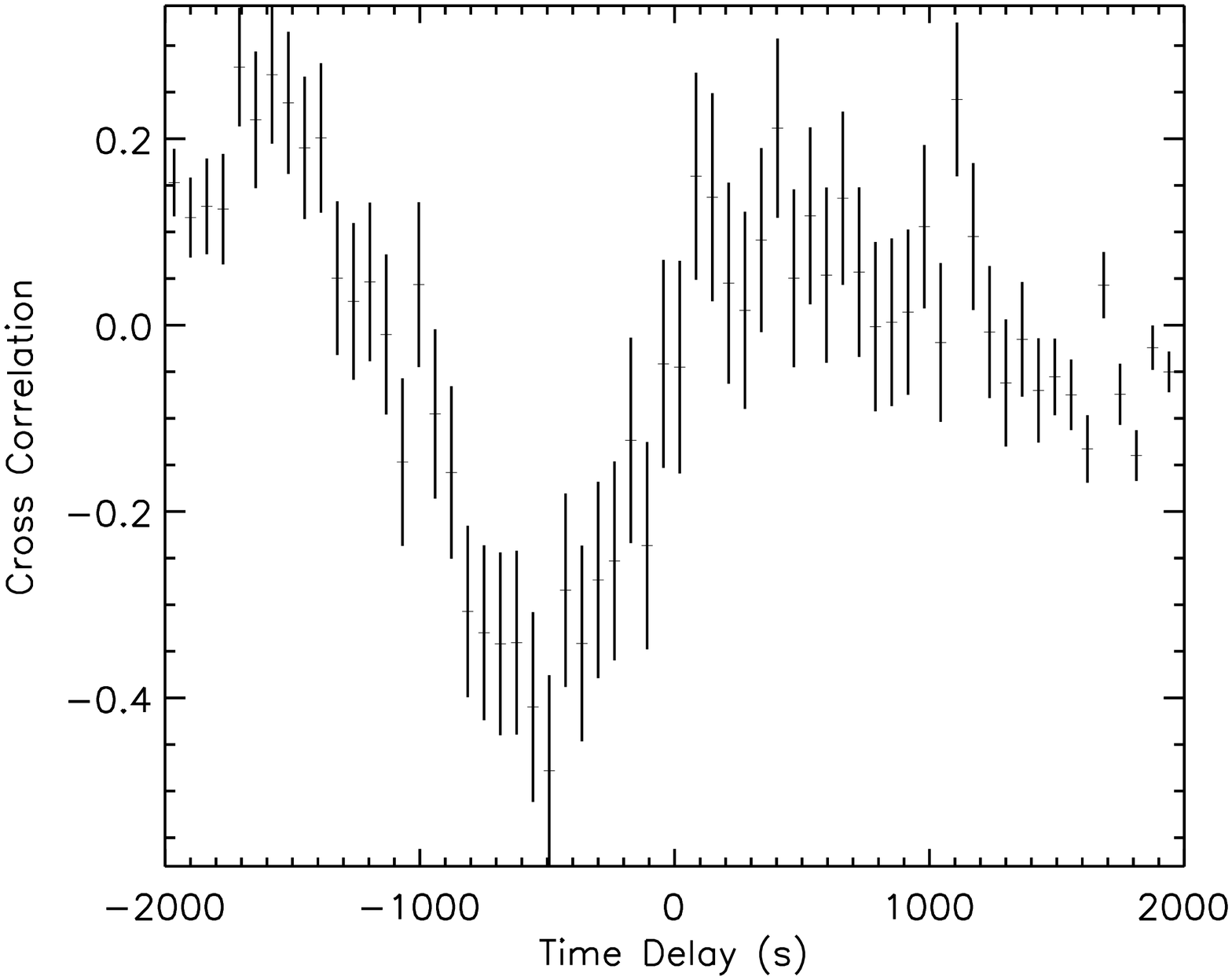}
\caption{Left: the CCFs of observation ID (ObsID) 40124-01-61-01 with time bin of 32 s, and the derived values of CCC and time lag are -0.28$\pm$0.05 and -501$\pm$63 s, respectively. Right: the CCFs of ObsID 40124-01-61-01 with time bin of 64 s, and the derived values of CCC and time lag are -0.37$\pm$0.03 and -498$\pm$57 s, respectively.}\label{fig:CCFs with 32 and 64 s time bin}
\end{figure*}

HEASOFT v 6.17, XSPEC v 12.9, and XRONOS v 5.22 are used for data reduction. Only the PCA Standard 2 form of data are used with 32 s bin size. The GTI files are created based on the criteria that the earth elevation angle is $>$ $10^{\circ}$ and the spacecraft pointing offset is $<$ $0.02^{\circ}$.
For the observation/segment whose X-ray intensity is greater than 40 counts s$^{-1}$ PCU$^{-1}$, the bright background model (pca$_{-}$bkgd$_{-}$cmbrightvle$_{-}$eMv20051128.mdl) is used, and for the others, the faint background model (pca$_{-}$bkgd$_{-}$cmfaintl7$_{-}$eMv20051128.mdl) is used.
We then use PCABACKEST v 4.0 to create PCA background files for extracting the background light curves.

\begin{figure*}
\centering
\includegraphics[width=5.6cm, height=4.7cm]{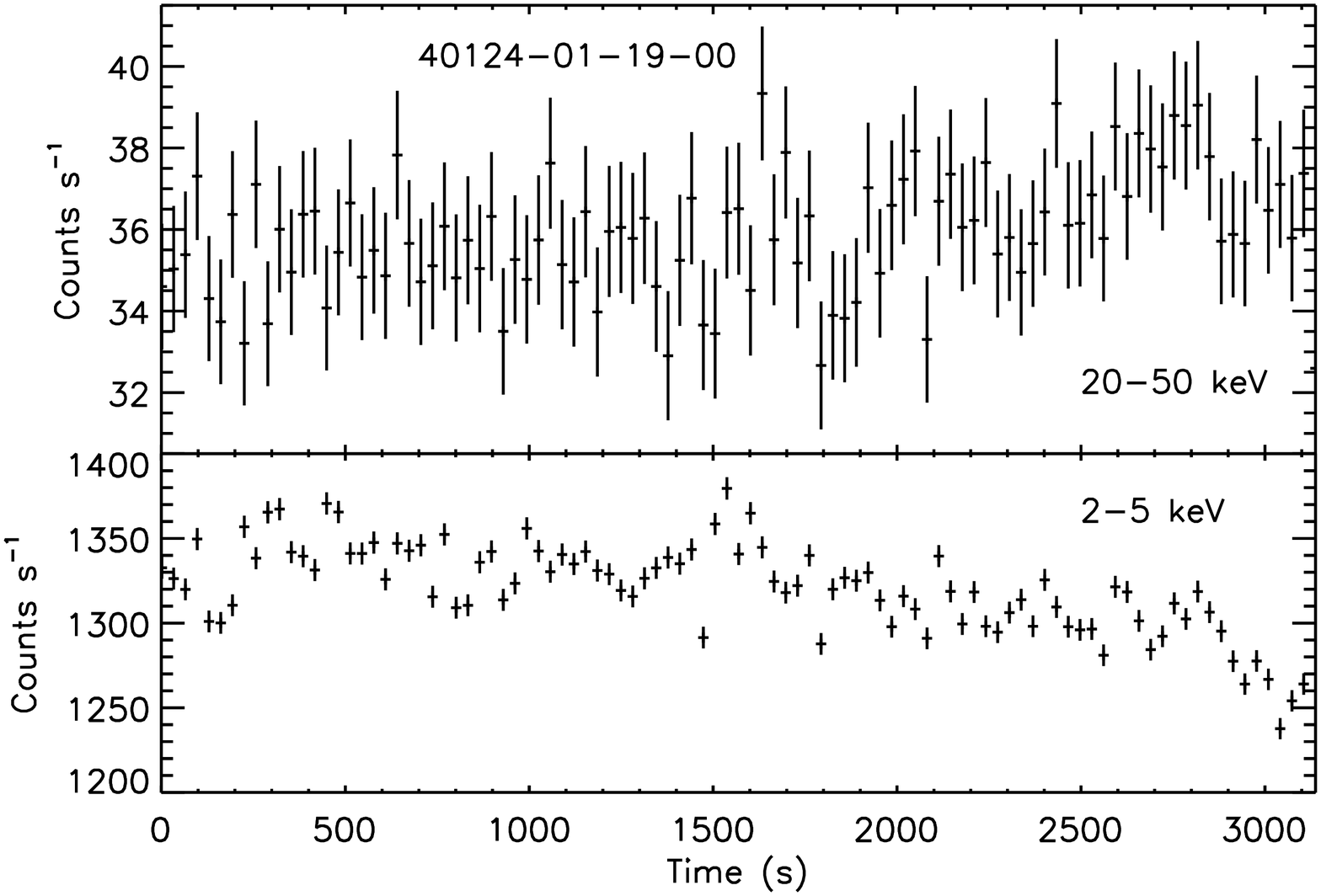}
\hspace{-0.62em}
\includegraphics[width=5.6cm, height=4.7cm]{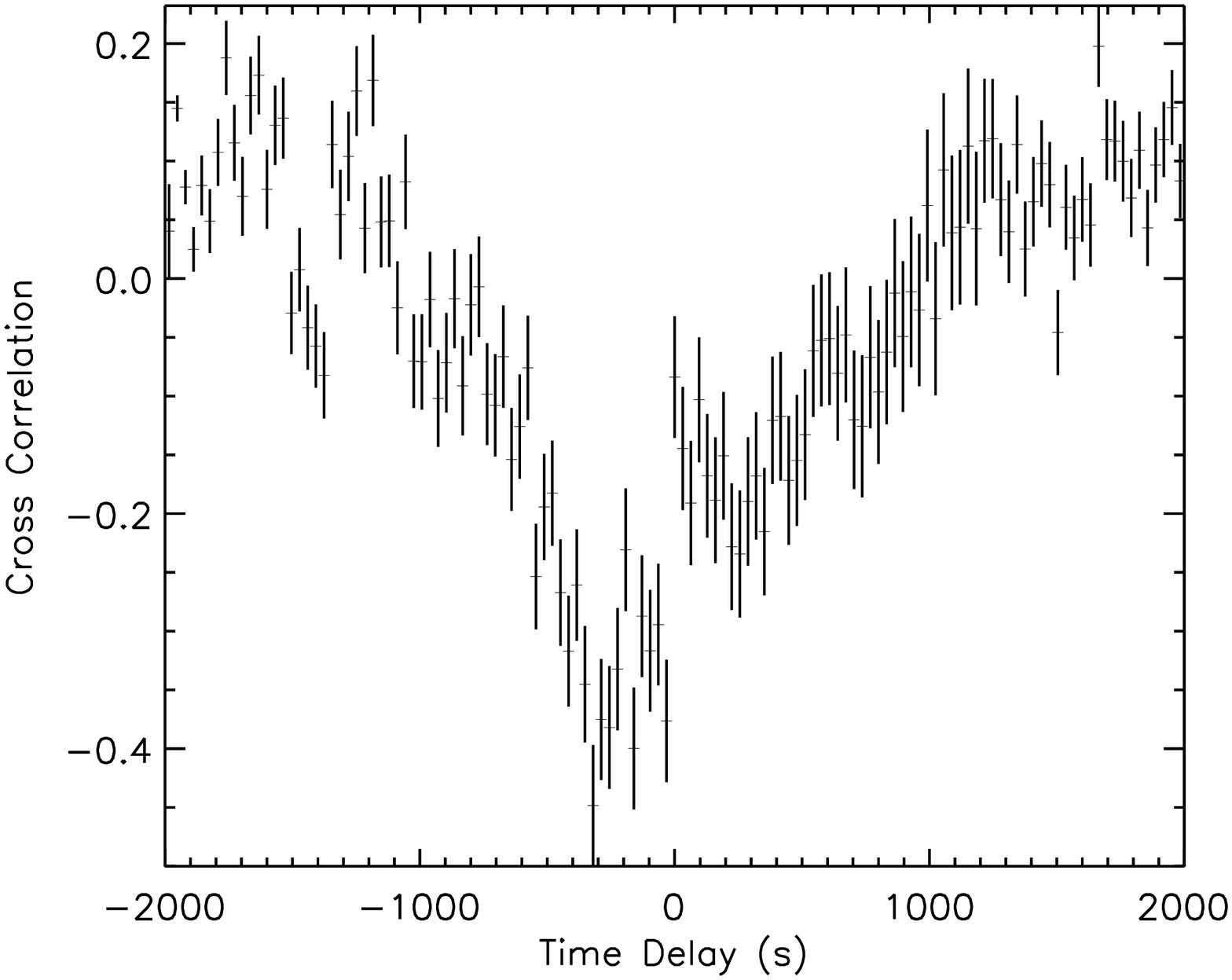}
\hspace{-0.62em}
\includegraphics[width=5.7cm, height=4.8cm]{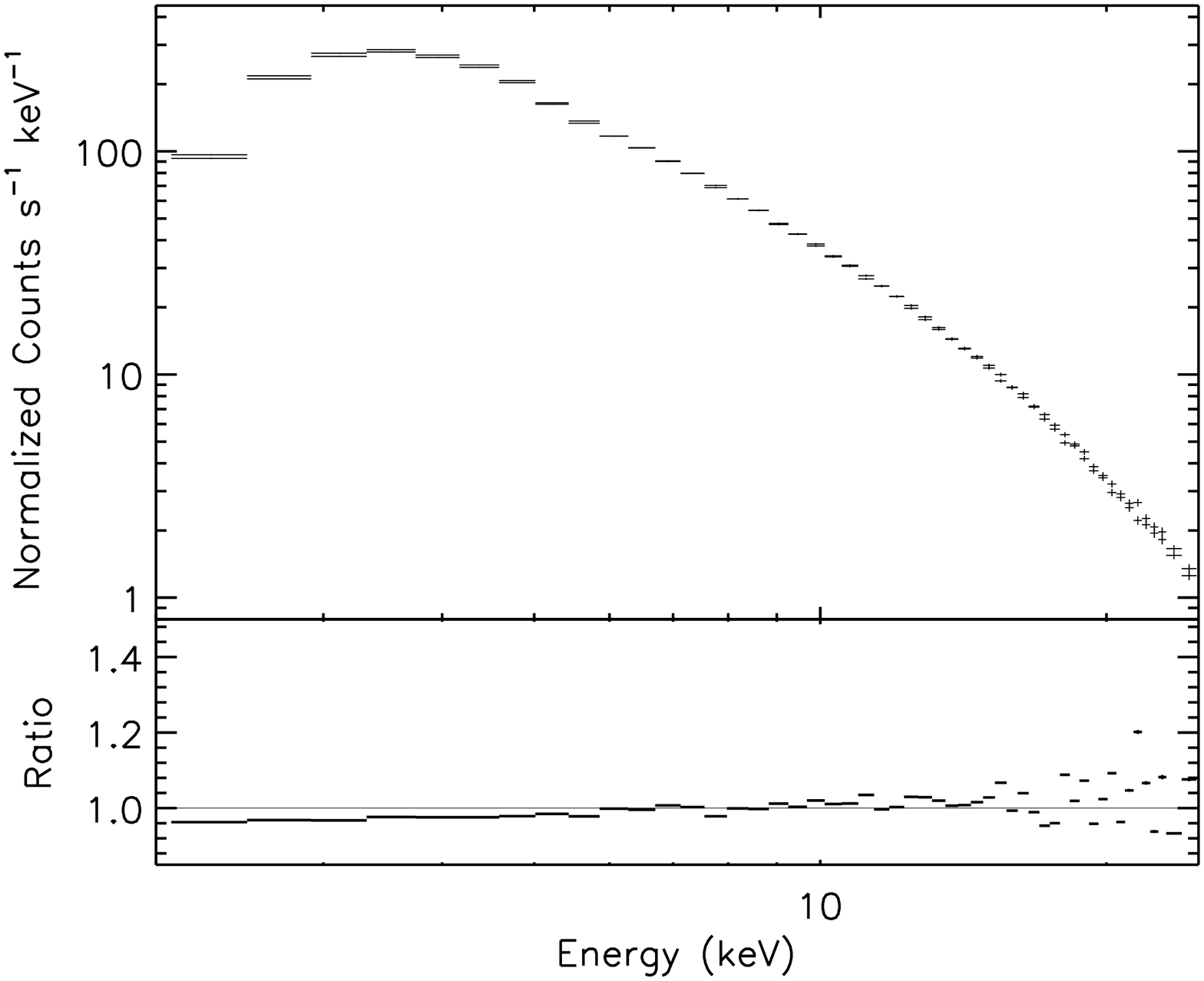}
\hspace{-0.62em}
\includegraphics[width=5.6cm, height=4.7cm]{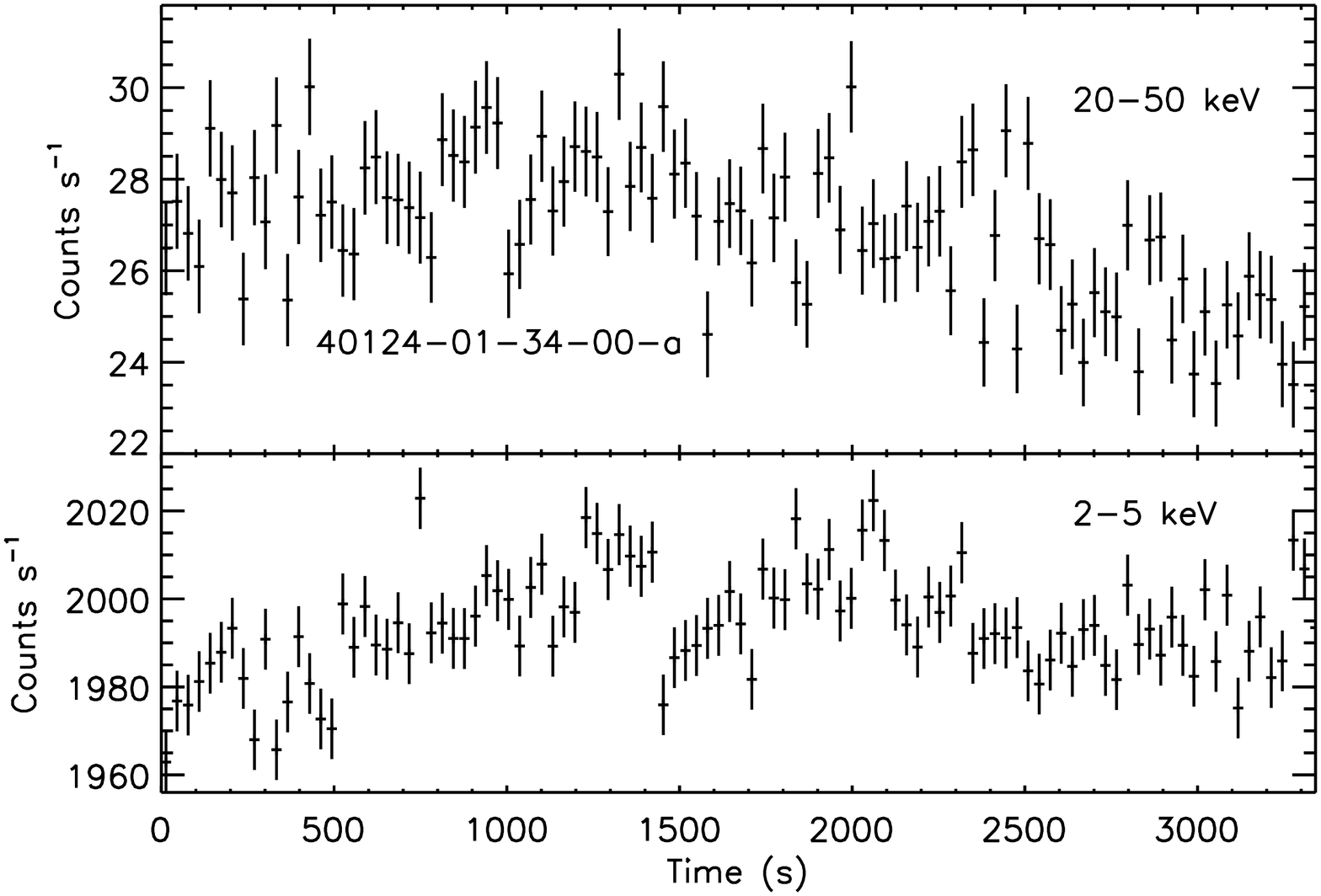}
\hspace{-0.62em}
\includegraphics[width=5.6cm, height=4.7cm]{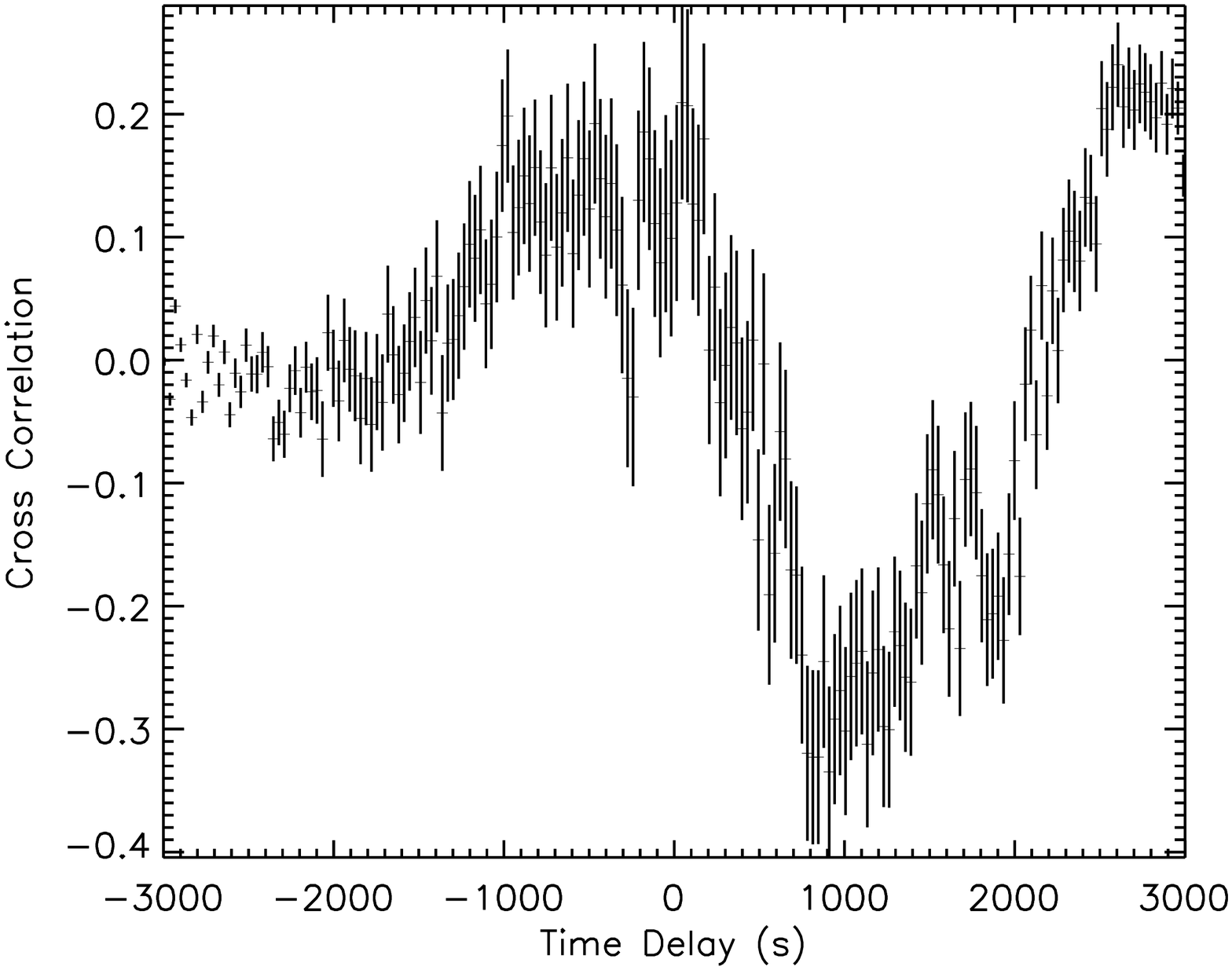}
\hspace{-0.62em}
\includegraphics[width=5.7cm, height=4.8cm]{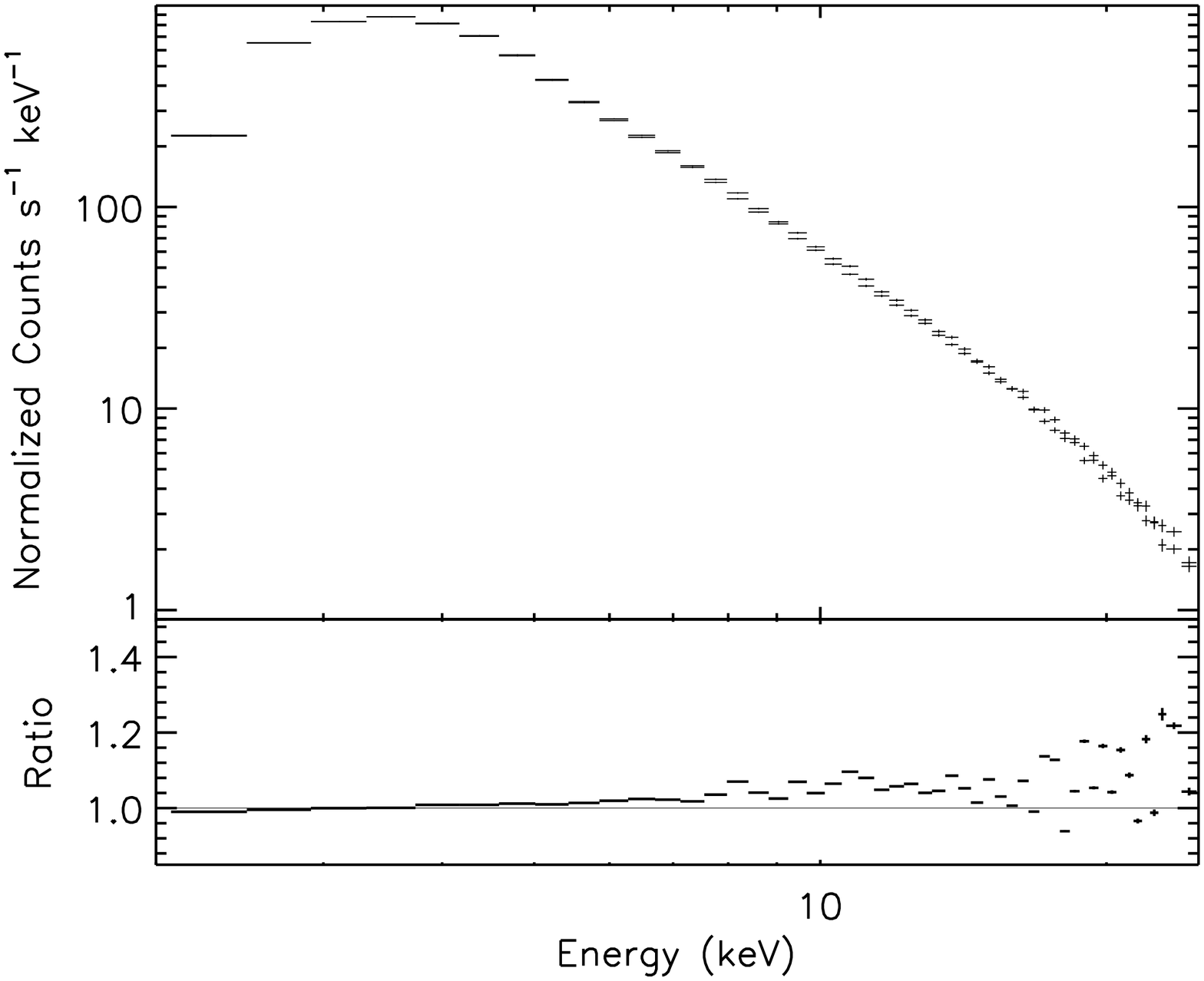}
\caption{Left: the PCA hard light curves (20-50 keV) and soft light curves (2-5 keV) for which anti-correlated soft lag (upper panel) and hard lag (lower panel) are detected. Middle: the corresponding cross-correlation functions. Right: the hard region spectra and soft region spectra and their ratio, which show spectral pivoting.}\label{fig:anti-correlated soft lag and hard lag}
\end{figure*}

PCU0 and PCU2 are used for the cross-correlation and spectra analyses as they are the only PCUs that always work for the 114 observations. CCF is then generated with the XRONOS program ``{\it crosscor}'' using the soft (2$-$5 keV) and hard (20$-$50 keV) \citep{2007ApJ...661.1055S} energy bands background-subtracted light curves. Following \citet{2007ApJ...661.1055S} and  \citet{2012ApJS..200...16S}, we fit the peak of CCFs with an Gaussian function to obtain the CCCs and time lags at 90 percent confidence level where the errors are obtained based on the criterion of $\Delta \chi^{2}$ = 2.7. Following \citet{2008ApJ...677..461L, 2013AJ....146...60L, 2014AJ....147...67L}, the background-subtracted spectra from the soft regions (the hardness ratio is 10\% less than the average at 20$-$50 keV/2$-$5 keV) and from the hard regions (the hardness ratio is 10\% more than the average at 20$-$50 keV/2$-$5 keV) are used to study spectral change.

\begin{figure*}
\centering
\includegraphics[width=5.6cm, height=4.7cm]{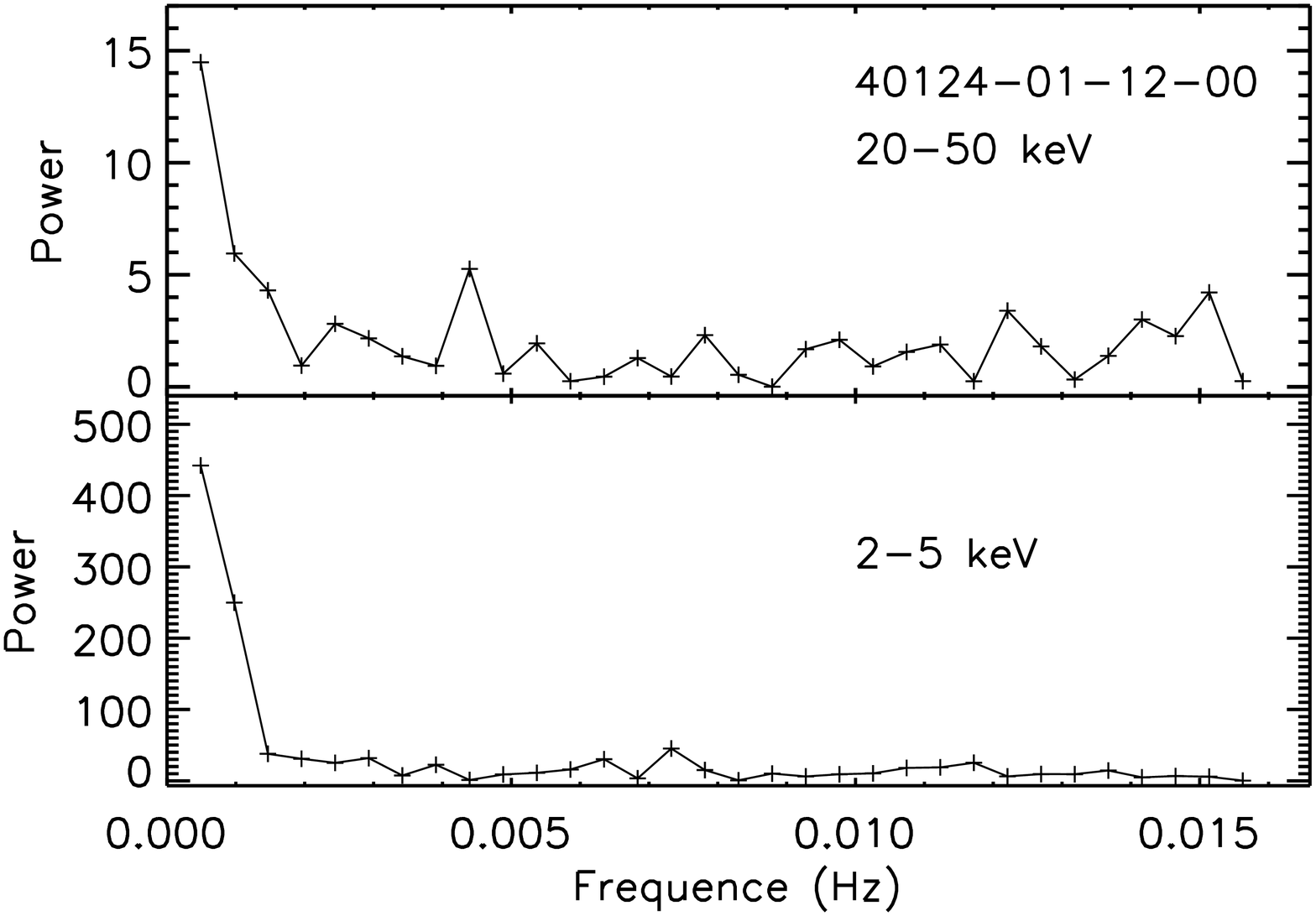}
\hspace{-0.62em}
\includegraphics[width=5.6cm, height=4.7cm]{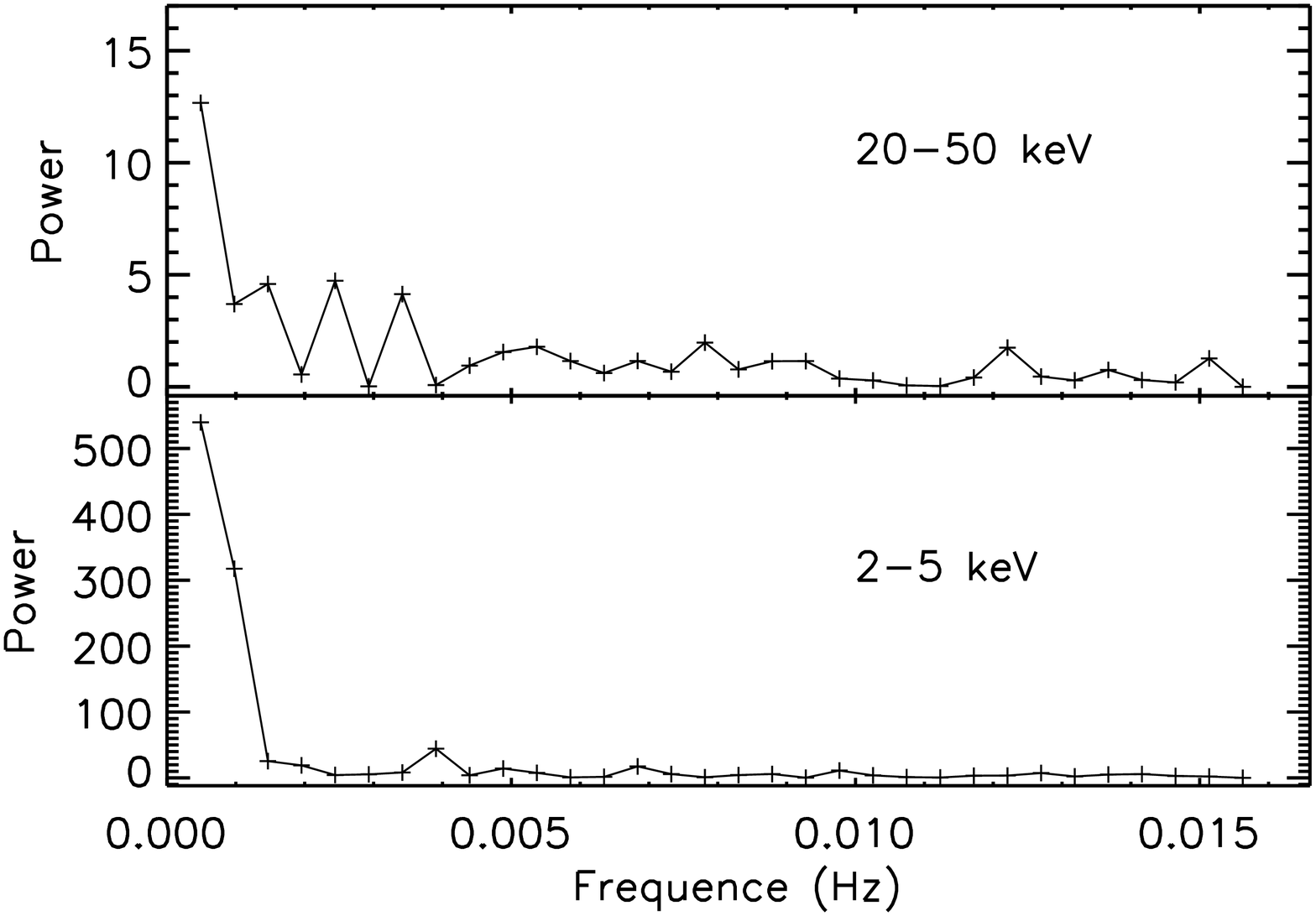}
\hspace{-0.62em}
\includegraphics[width=5.6cm, height=4.8cm]{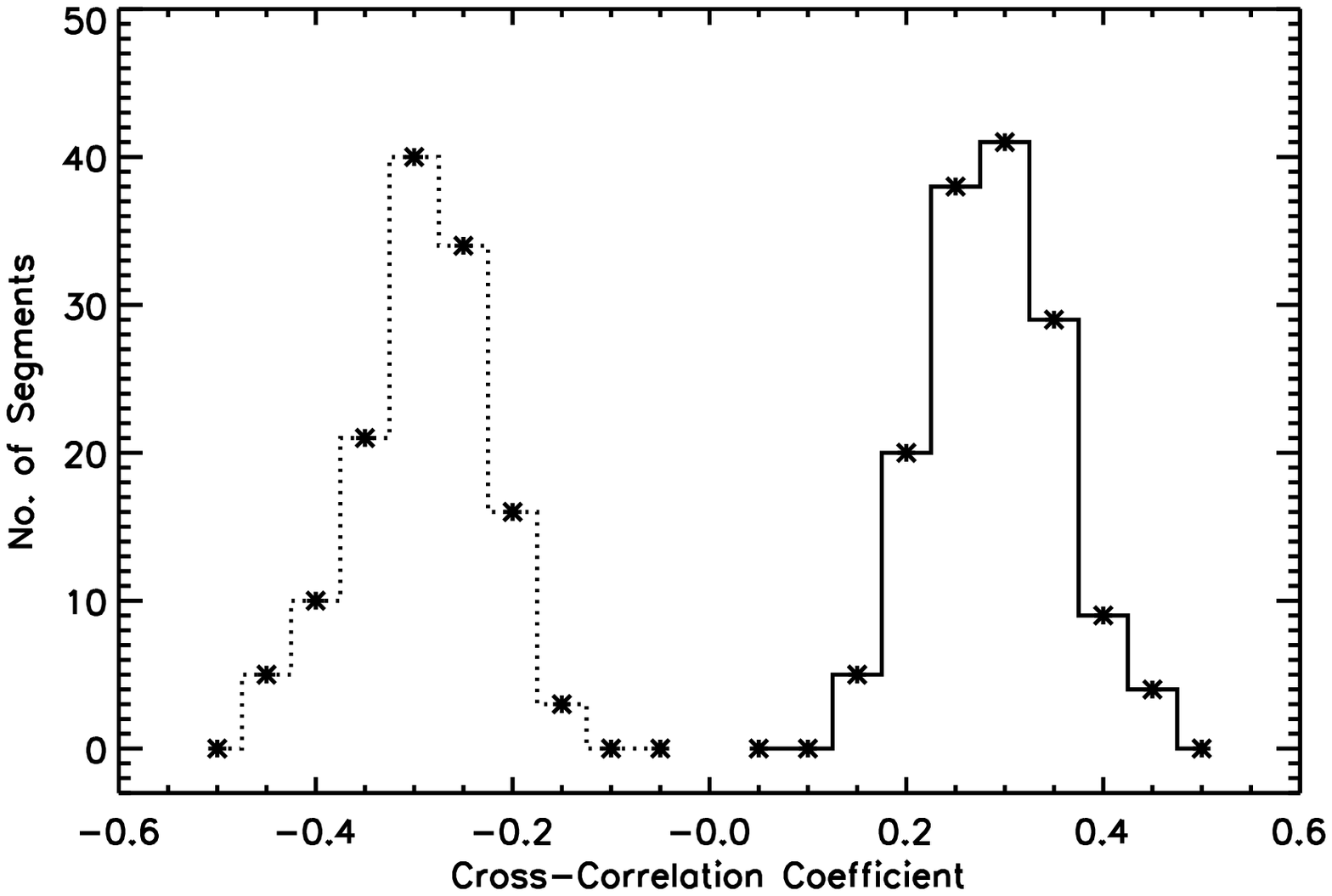}
\caption{Left: the PDS of hard light curve (20-50 keV) and soft light curve (2-5 keV) of 40124-01-12-00. Middle: the PDS of the two simulated light curves (the same duration as 40124-01-12-00). Right: histogram of the CCCs obtained from the simulated light curves. The CCCs of the anti-correlations (dotted line) and positive correlations (solid line) vs. the number of segments.}\label{fig: Distribution of CCCs from simulated light curves}
\end{figure*}

To check whether the values of time lags and CCCs change if the time bin changes, we also perform cross-correlation analysis on the PCA Standard 2 form of data with time bin of 64 s besides 32 s bin size. It turns out that the time lags do not change while the values of CCCs get a little larger when the time bin size is changed from 32 s to 64 s.
Fig.~\ref{fig:CCFs with 32 and 64 s time bin} shows a representative example of the CCFs with the 32 s and 64 s bin size of an observation.
To be consistent with other works, the time lags and CCCs are obtained from 32 s bin size of PCA Standard 2 form of data in this work.

\begin{figure*}
\centering
\includegraphics[width=5.6cm, height=4.7cm]{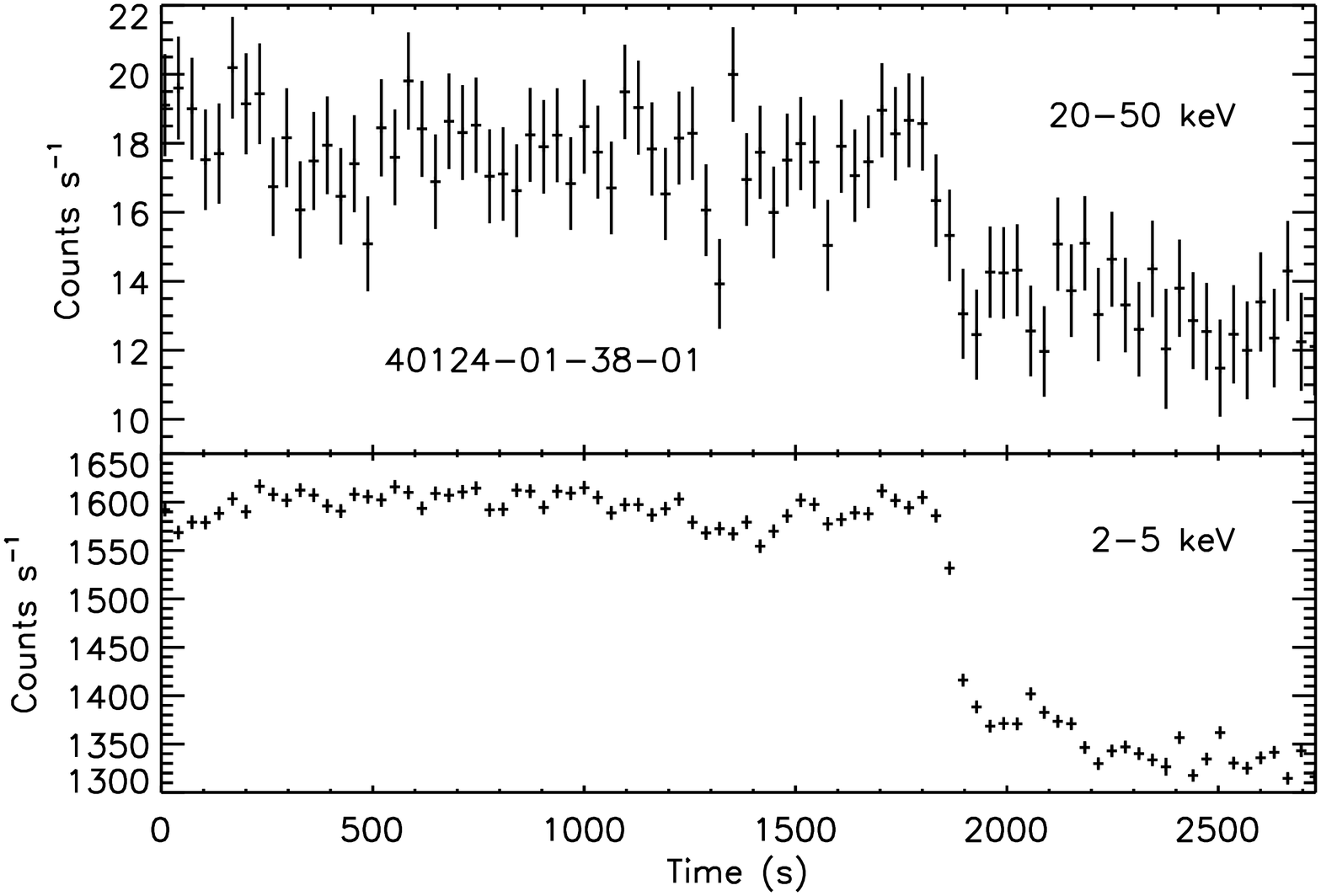}
\hspace{-0.62em}
\includegraphics[width=5.6cm, height=4.7cm]{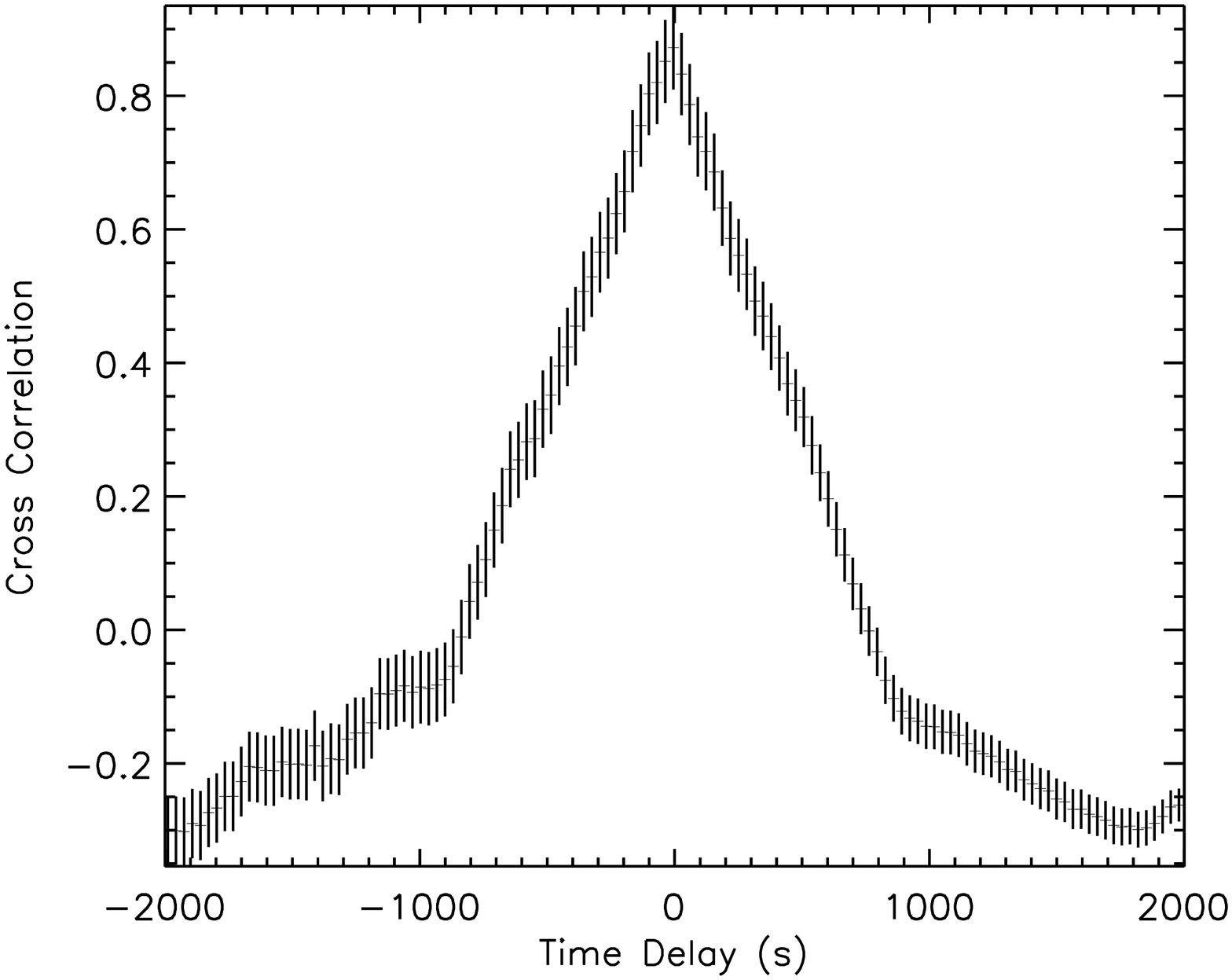}
\hspace{-0.62em}
\includegraphics[width=5.7cm, height=4.8cm]{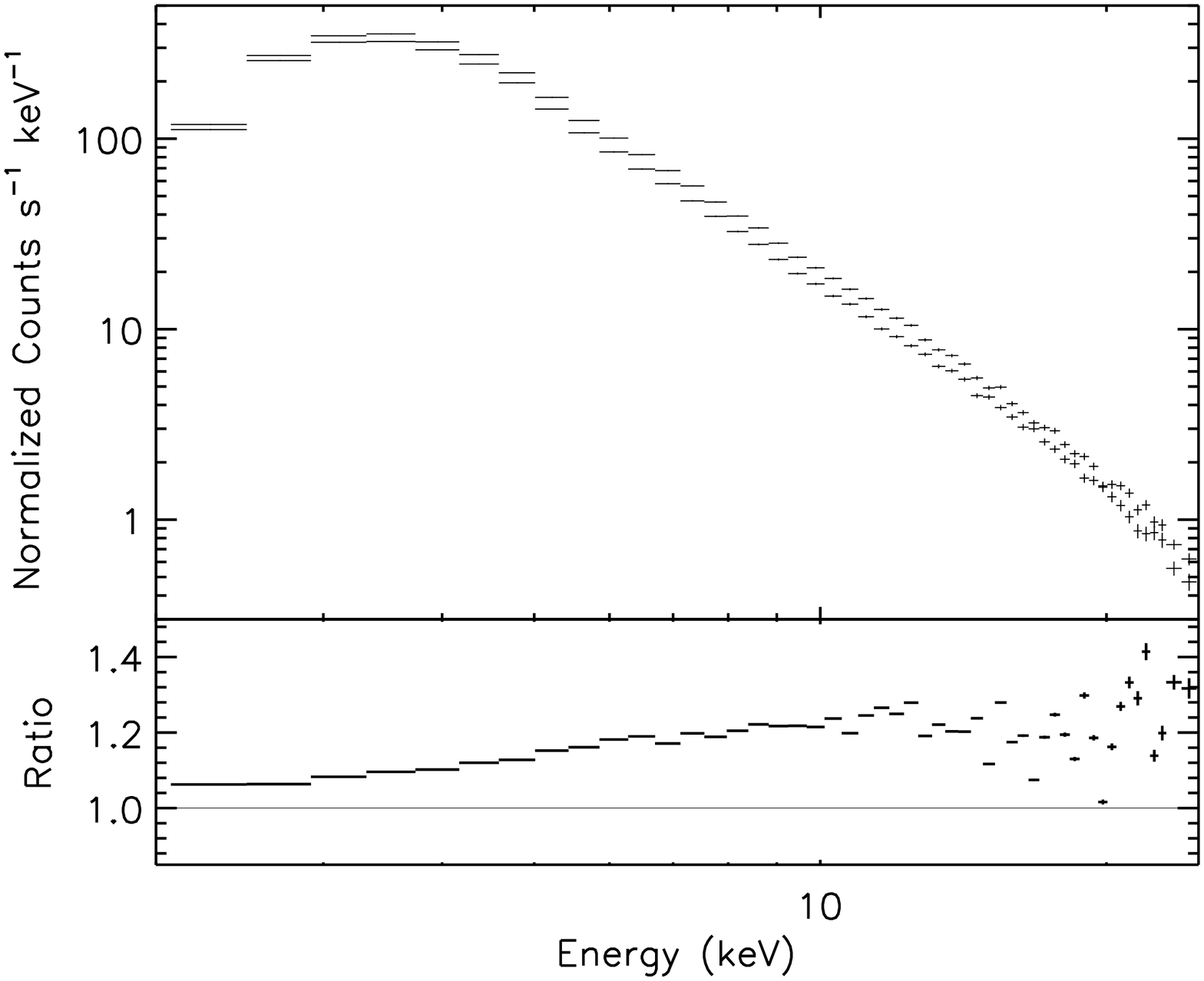}
\hspace{-0.62em}
\includegraphics[width=5.6cm, height=4.7cm]{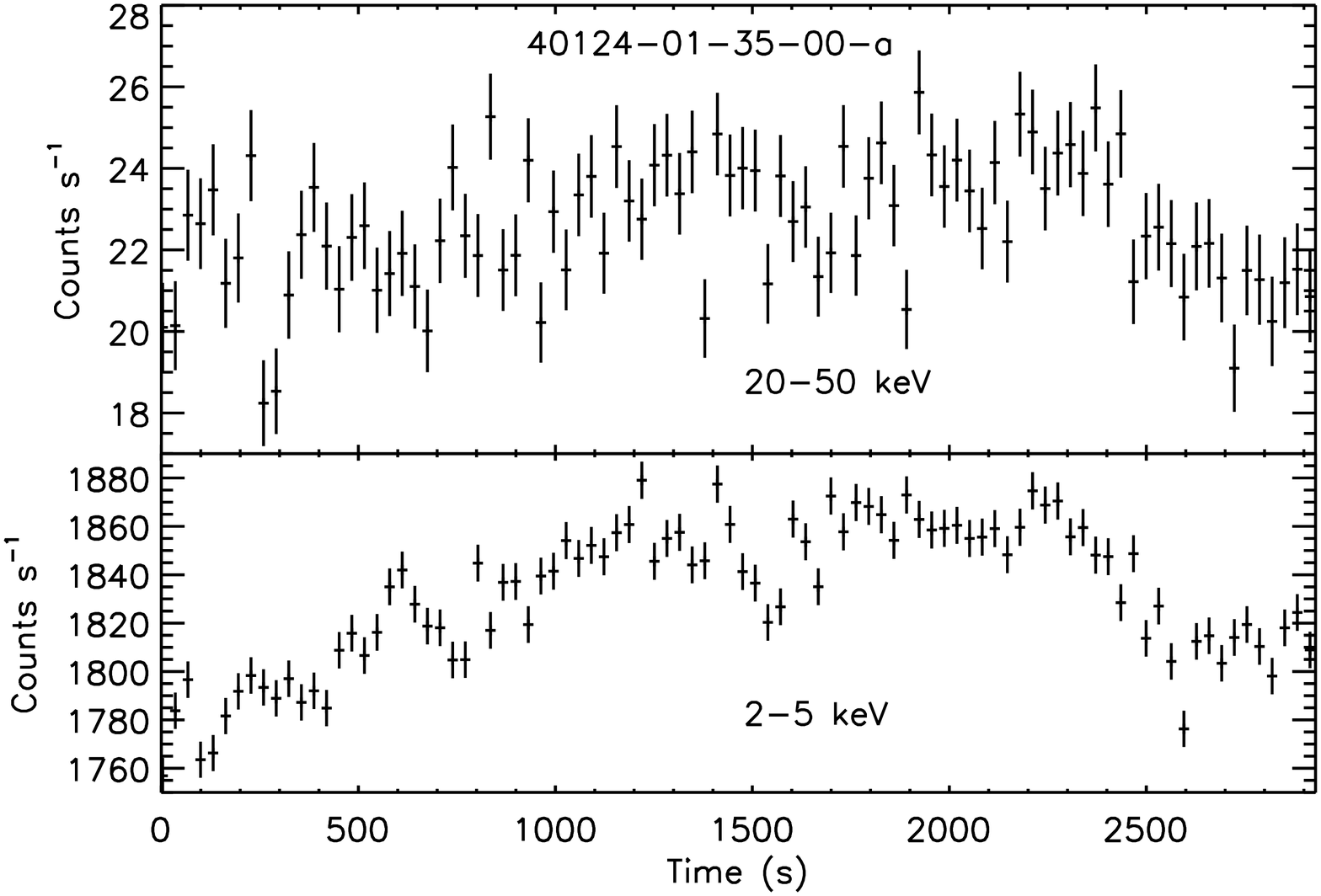}
\hspace{-0.62em}
\includegraphics[width=5.6cm, height=4.7cm]{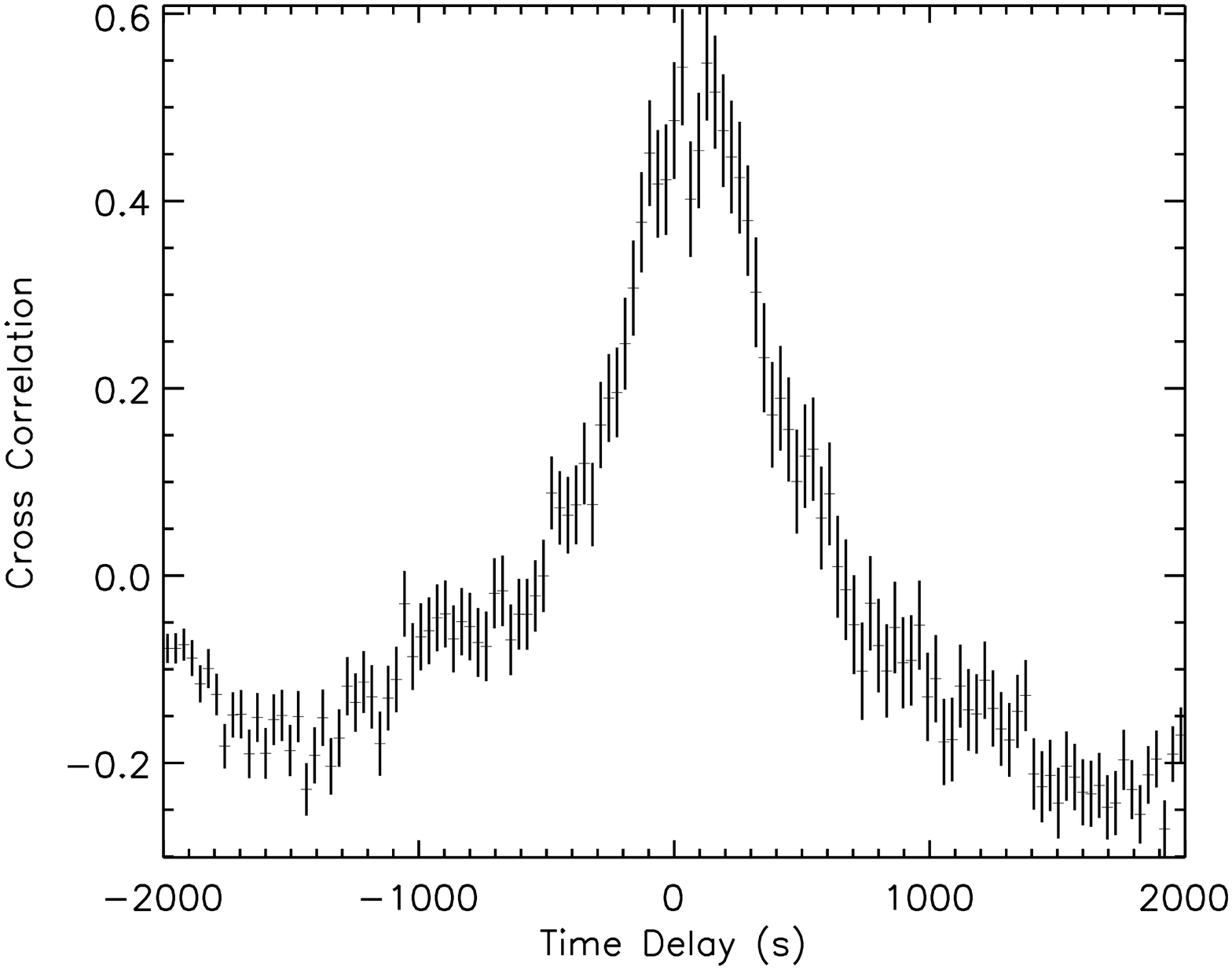}
\hspace{-0.62em}
\includegraphics[width=5.7cm, height=4.8cm]{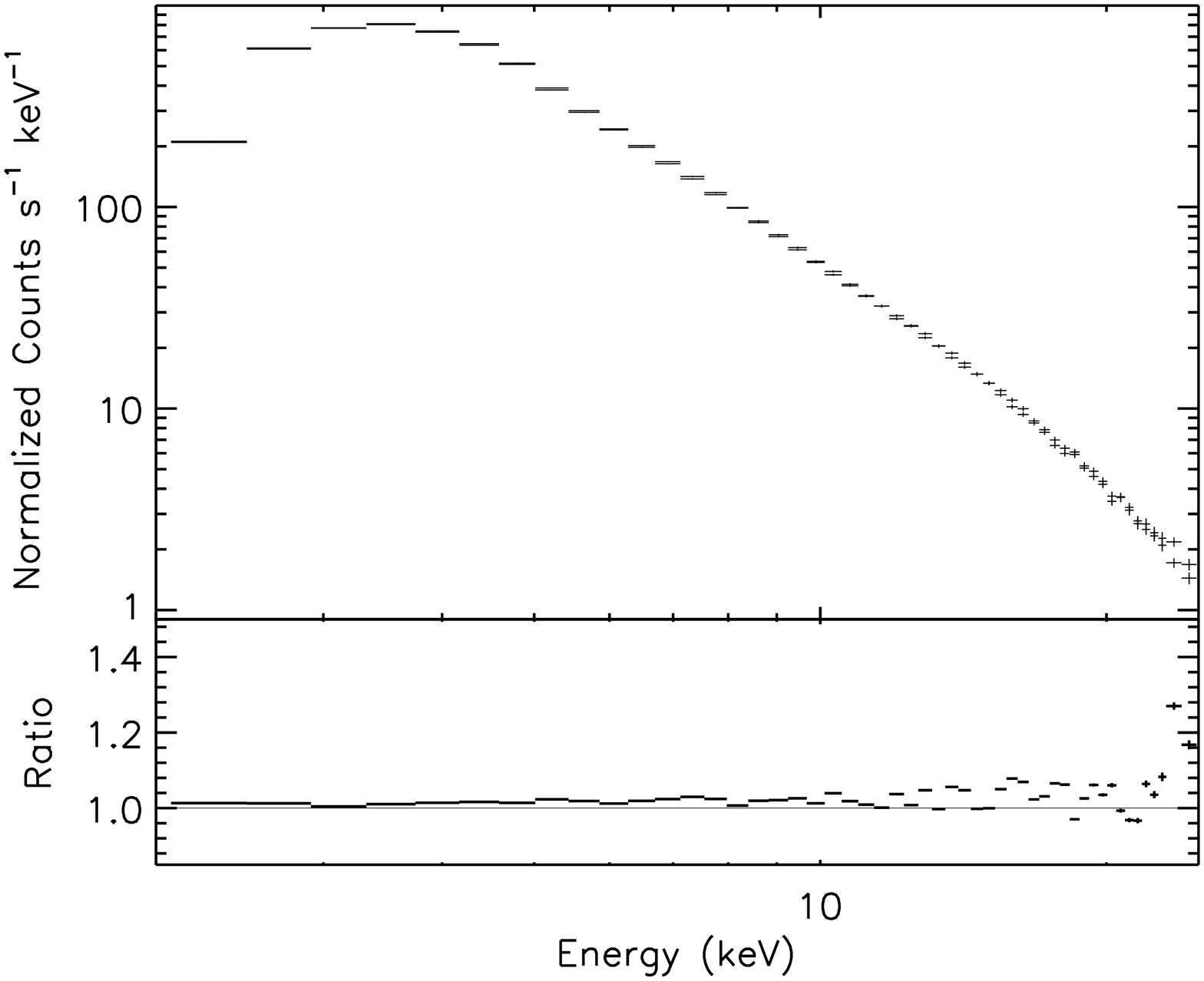}
\caption{Left: the PCA hard light curves (20-50 keV) and soft light curves (2-5 keV) for which positive correlated soft lag (upper panel) and hard lag (lower panel) are detected. Middle: the corresponding cross-correlation functions. Right: the hard region spectra and soft region spectra and their ratio, which do not show spectral pivoting.}\label{fig:positive correlated soft lag and hard lag}
\end{figure*}

\begin{figure*}
\centering
\includegraphics[width=5.6cm, height=4.7cm]{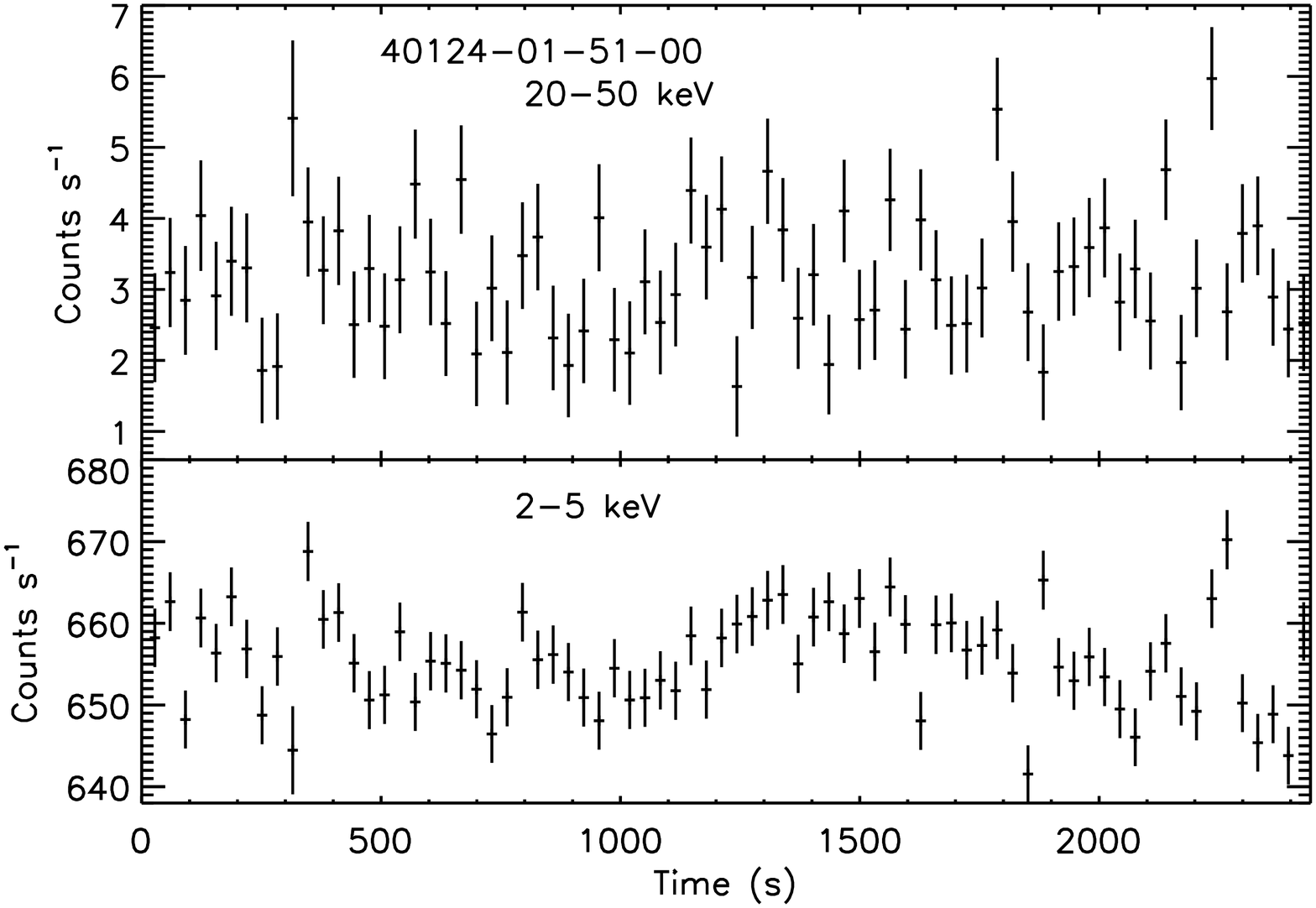}
\hspace{-0.62em}
\includegraphics[width=5.6cm, height=4.7cm]{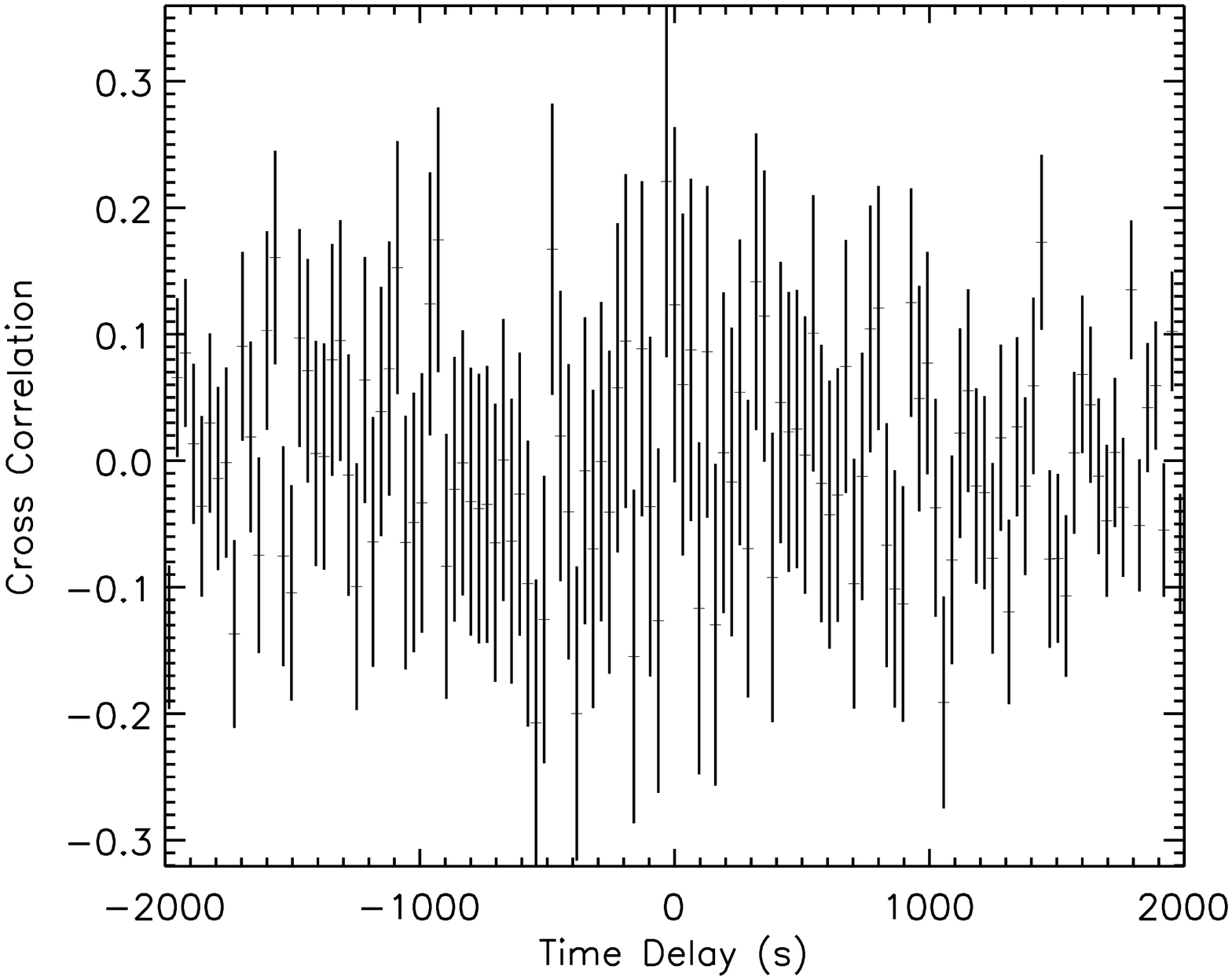}
\hspace{-0.62em}
\includegraphics[width=5.7cm, height=4.8cm]{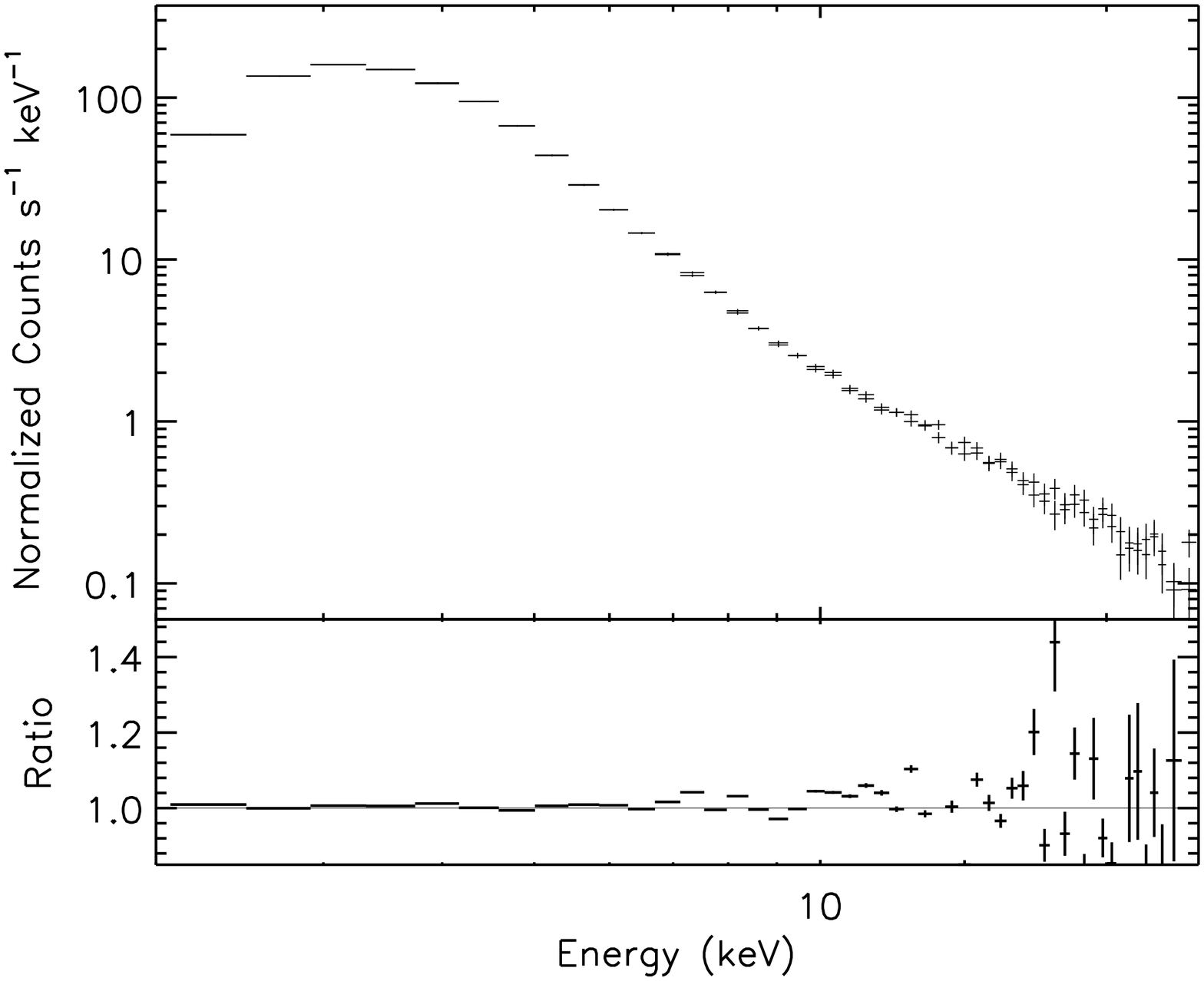}

\caption{Left: the PCA hard light curves (20-50 keV) and soft light curves (2-5 keV) showing a typical ambiguous correlation. Middle: the corresponding cross-correlation functions. Right: the hard region spectra and soft region spectra and their ratio, which do not show spectral pivoting.}\label{fig:ambiguous correlation}
\end{figure*}

Light curve and HID are used for study of the evolution of cross-correlations and time lags. The intensity is defined as the background subtracted light curve of 2$-$20 keV, the hardness is defined as the count-rate ratio of 6$-$20 keV/2$-$6 keV, and only data of PCU2 are used, which are the same as given by \citet{2014AdSpR..54.1678R}.

\section[]{RESULTS}

After performing cross-correlation analysis on all the 140 observations/segments during the 1999$-$2000 outburst of XTE J1859+226, both anti-correlations and positive correlations with soft or hard X-ray long time-scale lags are detected.

\begin{figure*}
\centering
\includegraphics[width=17.0cm]{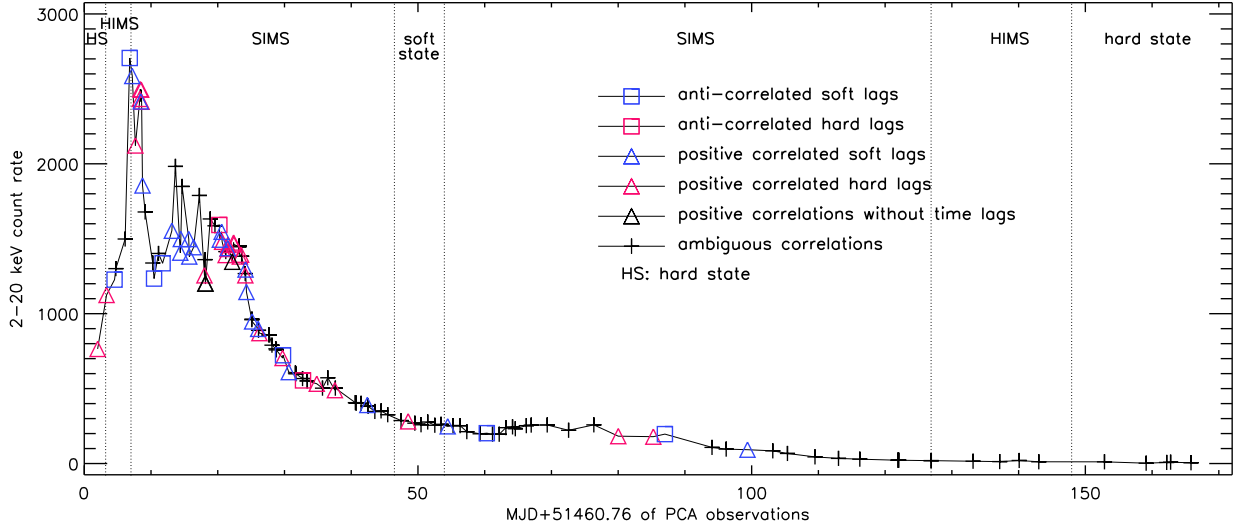}

\caption{Plot showing cross-correlation and time lag from the PCA observation for anti-correlated soft lag, anti-correlated hard lag, positive correlated soft lag, positive correlated hard lag, positive correlation without time lag and ambiguous correlation segments, as shown in the legend.}\label{fig:lc}
\end{figure*}

\begin{figure}
\centering
\includegraphics[width=8.3cm, height=6.15cm]{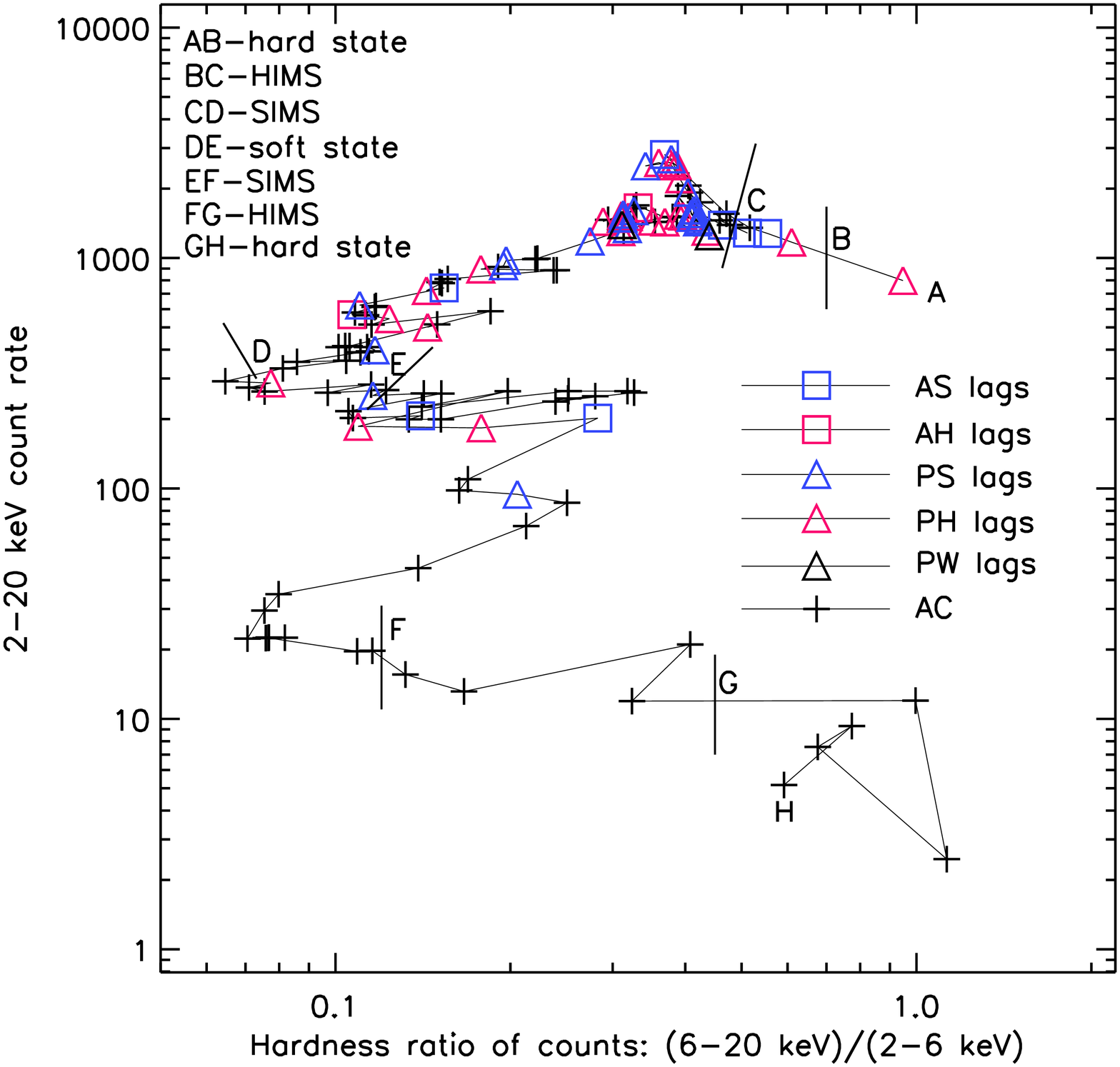}

\caption{The results of cross-correlation and time lag from the PCA observation, and the anti-correlated soft (AS) lag, anti-correlated hard (AH) lag, positive correlated soft (PS) lag, positive correlated hard (PH) lag, positive correlation without (PW) time lag and ambiguous correlation (AC) segments, as shown in the legend.}\label{fig:HID}
\end{figure}

9 out of 140 observations/segments display anti-correlations in HIMS and SIMS as shown in Table~\ref{tab:anti-correlation}.
Among the observations with anti-correlations, soft lags and hard lags are detected in 7 and 2 observations/segments, respectively, where soft lags are located in HIMS and SIMS, and hard lags are located in SIMS (see Table~\ref{tab:anti-correlation}).
Clearly, the number of the observations/segments with anti-correlation soft lags is more than that with anti-correlation hard lags.
Fig.~\ref{fig:anti-correlated soft lag and hard lag} shows a representative example of anti-correlated soft lag (upper panel) and hard lag (lower panel), in which the observations with anti-correlation show spectral pivoting. Spectral pivoting means that the spectral of soft regions and the spectral of hard regions intersect at a pivoting point which corresponds to the point of unit spectral ratio \citep{2003A&A...397..711R, 2003ApJ...593..452C}.

\begin{table*}
\caption{Log of all observations ($>$ 1000 s) in which anti-correlations are detected.}
\begin{center}
\renewcommand{\arraystretch}{1.0}
\label{tab:anti-correlation}
\begin{tabular}{l c c c c c}
\hline \hline
      ObsID& Start Time& Stop Time& State& CCC& Time Lag (s)  \\
\hline

\multicolumn{6}{c}{Hard lags}\\

\hline
40124-01-34-00-a&1999-10-30, 00:19&01:17&SIMS$^{(1)}$&-0.31$\pm$0.04&953$\pm$65\\
40124-01-47-00-b&1999-11-11, 12:54&13:34&SIMS&-0.30$\pm$0.05& 204$\pm$87\\

\hline

\multicolumn{6}{c}{Soft lags}\\

\hline

40124-01-08-00&1999-10-14, 07:19&08:10&HIMS$^{(2)}$&-0.27$\pm$0.06&-143$\pm$36\\
40124-01-12-00&1999-10-16, 13:57&14:30&HIMS&-0.42$\pm$0.04&-635$\pm$29\\
40124-01-19-00&1999-10-20, 05:22&06:17&SIMS&-0.39$\pm$0.03&-276$\pm$32\\
40124-01-21-00&1999-10-21, 12:04&12:43&SIMS&-0.31$\pm$0.04&-120$\pm$16\\
40124-01-44-00-b&1999-11-08, 13:00&13:42&SIMS&-0.25$\pm$0.06&-718$\pm$75\\
40124-01-56-00&1999-12-09, 01:58&02:45&SIMS&-0.15$\pm$0.03&-124$\pm$70\\
40124-01-61-01&2000-01-04, 18:24&19:17&SIMS&-0.28$\pm$0.05&-501$\pm$63\\
\hline \hline

\end{tabular}
\end{center}
\footnotesize{
\begin{quote}{\bf Notes.}
The date, state of the detected anti-correlations on the HID, cross-correlation coefficient and time lag of each observation are listed. We label the segments ($>$ 1000 s) of an observation by `a' , `b', and so on.

$^{(1)}$The total rms of the PDS ranges from 7\% to 14\%, the photon index ranges from 1.9 to 3.2, the disk temperature ranges from 0.70 keV to 0.85 keV, and the fold energy ranges from 108.3 keV to 201.2 keV \citep{2014AdSpR..54.1678R}.

$^{(2)}$The total rms of the PDS ranges from 7\% to 24\%, the QPO frequency ranges from 3.05 Hz to 5.97 Hz, the photon index ranges from 2.0 to 2.4, the disk temperature ranges from 0.75 keV to 0.85 keV, and the fold energy ranges from 64 keV to 180 keV \citep{2014AdSpR..54.1678R}.

\end{quote}
}
\end{table*}

\begin{table*}
\small
\caption{Log of all observations ($>$ 1000 s) in which positive correlations are detected.}
\begin{center}
\renewcommand{\arraystretch}{1.0}
\label{tab:positive correlations}
\begin{tabular}{l c c c c c}
\hline \hline
     ObsID& Start Time& Stop Time& State& CCC& Time Lag (s)  \\
\hline

\multicolumn{6}{c}{Hard lags}\\

\hline

40124-01-04-00&1999-10-11, 18:25&18:44&hard state$^{(1)}$&0.43$\pm$0.09&170.8$\pm$48.2\\
40124-01-06-00&1999-10-13, 02:37&03:50&HIMS&0.26$\pm$0.05&87.3$\pm$42.6\\
40124-01-14-00&1999-10-17, 10:11&11:08&SIMS&0.76$\pm$0.04&12.1$\pm$3.7\\
40122-01-01-03&1999-10-18, 02:13&02:57&SIMS&0.39$\pm$0.06&13.7$\pm$7.5\\
40122-01-01-02&1999-10-18, 03:49&04:28&SIMS&0.35$\pm$0.04&214.7$\pm$46.5\\
40122-01-01-00-a&1999-10-18, 07:01&07:59&SIMS&0.65$\pm$0.03&23.3$\pm$15.2\\
40122-01-01-00-b&1999-10-18, 08:36&09:34&SIMS&0.43$\pm$0.05&25.7$\pm$16.5\\
40124-01-31-00&1999-10-27, 18:39&19:27&SIMS&0.42$\pm$0.04&37.3$\pm$19.5\\
40124-01-34-02&1999-10-30, 10:08&10:48&SIMS&0.46$\pm$0.03&38.6$\pm$24.8\\
40124-01-35-01-a&1999-10-30, 22:41&23:07&SIMS&0.28$\pm$0.04&107.3$\pm$19.2\\
40124-01-35-00-a&1999-10-31, 05:11&06:02&SIMS&0.52$\pm$0.04& 80.5$\pm$16.3\\
40124-01-35-00-c&1999-10-31, 08:22&09:11&SIMS&0.34$\pm$0.03&262.5$\pm$41.7\\
40124-01-36-00-a&1999-11-01, 03:26&04:23&SIMS&0.33$\pm$0.03&273.8$\pm$50.8\\
40124-01-36-00-b&1999-11-01, 05:02&05:59&SIMS&0.25$\pm$0.03&105.3$\pm$24.6\\
40124-01-37-01&1999-11-02, 01:50&02:25&SIMS&0.36$\pm$0.04&241.4$\pm$80.7\\
40124-01-37-02-a&1999-11-02, 06:37&07:32&SIMS&0.23$\pm$0.02&327.4$\pm$54.6\\
40124-01-38-00-b&1999-11-02, 21:26&21:51&SIMS&0.34$\pm$0.04& 23.9$\pm$20.4\\
40124-01-41-00&1999-11-05, 00:13&01:20&SIMS&0.60$\pm$0.02& 30.7$\pm$21.0\\
40124-01-44-00-a&1999-11-08, 11:24&12:05&SIMS&0.31$\pm$0.03&157.8$\pm$36.7\\
40124-01-48-01&1999-11-13, 14:25&15:01&SIMS&0.19$\pm$0.02&67.2$\pm$40.9\\
40124-01-50-00&1999-11-16, 07:42&08:19&SIMS&0.37$\pm$0.03&105.5$\pm$20.5\\
40124-01-53-01&1999-11-27, 07:12&07:53&soft state$^{(2)}$&0.30$\pm$0.03&165.4$\pm$33.0\\
40124-01-60-01&1999-12-28, 18:44&19:39&SIMS&0.27$\pm$0.03& 61.4$\pm$34.3\\
40124-01-61-00&2000-01-03, 00:52&02:29&SIMS&0.15$\pm$0.02&402.8$\pm$62.0\\

\hline

\multicolumn{6}{c}{Soft lags}\\

\hline

40124-01-13-00&1999-10-16, 23:04&23:34&SIMS&0.46$\pm$0.05&-83.6$\pm$29.1\\
40122-01-01-01&1999-10-18, 05:25&06:22&SIMS&0.31$\pm$0.04&-235.7$\pm$41.7\\
40124-01-16-00&1999-10-18, 11:50&12:40&SIMS&0.56$\pm$0.04&-27.4$\pm$21.6\\
40124-01-23-01&1999-10-22, 21:22&21:40&SIMS&0.50$\pm$0.08&-31.9$\pm$17.4\\
40124-01-26-00-a&1999-10-24, 03:42&04:40&SIMS&0.26$\pm$0.02&-361.6$\pm$40.5\\
40124-01-26-00-b&1999-10-24, 05:17&06:17&SIMS&0.36$\pm$0.04&-17.3$\pm$14.4\\
40124-01-28-00&1999-10-25, 10:17&11:04&SIMS&0.63$\pm$0.09&-19.4$\pm$6.9\\
40124-01-28-01&1999-10-25, 12:01&12:23&SIMS&0.36$\pm$0.05&-82.8$\pm$50.9\\
40124-01-29-00&1999-10-26, 03:38&04:49&SIMS&0.41$\pm$0.03&-217.8$\pm$31.6\\
40124-01-34-00-b&1999-10-30, 01:54&02:52&SIMS&0.53$\pm$0.04&-56.8$\pm$27.5\\
40124-01-34-01&1999-10-30, 08:26&09:06&SIMS&0.20$\pm$0.04&-60.7$\pm$26.9\\
40124-01-35-00-b&1999-10-31, 06:41&07:36&SIMS&0.39$\pm$0.03&-235.8$\pm$38.6\\
40124-01-38-00-c&1999-11-02, 22:31&23:29&SIMS&0.52$\pm$0.04&-60.4$\pm$22.6\\
40124-01-38-01&1999-11-03, 01:48&02:35&SIMS&0.84$\pm$0.02&-12.6$\pm$3.7\\
40124-01-39-00-c&1999-11-03, 22:30&23:29&SIMS&0.50$\pm$0.03&-133.9$\pm$32.7\\
40124-01-40-00&1999-11-04, 19:51&20:16&SIMS&0.31$\pm$0.09&-67.9$\pm$24.3\\
40124-01-45-00&1999-11-09, 09:43&10:30&SIMS&0.40$\pm$0.04&-168.5$\pm$34.9\\
40124-01-51-02-a&1999-11-21, 04:17&05:13&SIMS&0.16$\pm$0.02&-204.8$\pm$57.1\\
40124-01-55-00&1999-12-03, 05:20&05:59&SIMS&0.32$\pm$0.04&-34.7$\pm$26.9\\
40124-01-63-00&2000-01-17, 03:23&04:56&SIMS&0.21$\pm$0.02&-160.6$\pm$42.7\\

\hline

\multicolumn{6}{c}{Without time lags}\\

\hline

40122-01-02-00-b&1999-10-27, 21:00&21:22&SIMS&0.27$\pm$0.06&-13.6$\pm$29.7\\
40122-01-02-00-c&1999-10-27, 22:37&23:03&SIMS&0.26$\pm$0.04&-9.8$\pm$31.9\\
40124-01-36-01&1999-10-31, 22:42&23:06&SIMS&0.38$\pm$0.03&12.6$\pm$13.6\\

\hline \hline

\end{tabular}
\end{center}
\footnotesize{
\begin{quote}{\bf Notes.}
The date, state of the detected positive correlations on the HID, cross-correlation coefficient and time lag of each observation are listed. The segments ($>$ 1000 s) of an observation are labeled by `a' , `b', and so on.

$^{(1)}$The total rms of the PDS ranges from 24\% to 30\%, the QPO frequency ranges from 0.46 Hz to 1.19 Hz, the photon index ranges from 1.6 to 2.1, the disk temperature ranges from 0.83 keV to 1.01 keV, and the fold energy ranges from 53 keV to 114 keV \citep{2014AdSpR..54.1678R}.

$^{(2)}$The total rms of the PDS ranges from 1\% to 2\%, the photon index remains $\sim$ 2.2, the disk temperature ranges from 0.7 keV to 0.9 keV \citep{2014AdSpR..54.1678R}.

\end{quote}
}
\end{table*}

To confirm the detected anti-correlation, we follow \citet{2010ApJ...725.1317S, 2012ApJS..200...16S, 2014AJ....147...67L} by simulating two long independent light curves (2.24 $\times$ 10$^{6}$ s) using the method described in \citet{1995A&A...300..707T}. To obtain the two simulated light curves, we use the parameters of the observation ID (ObsID) 40124-01-12-00 where the highest anti-correlation coefficient ($\sim$--0.42) is detected.
For this observation, the mean count rate, standard deviation and power density spectrum (PDS) power-law index $\beta$ of the soft energy band light curve are 3253.7 counts s$^{-1}$, 43.2 counts s$^{-1}$ and --1.32, and the mean count rate, standard deviation and PDS power-law index $\beta$ of the hard energy band light curve are 44.5 counts s$^{-1}$, 1.8 counts s$^{-1}$ and --0.83, respectively.
To obtain the power-law index, a power-law model is used to fit the corresponding PDS. We perform cross-correlation analysis on these two simulated light curves with a bin size of 32 s and a segment length of 3200 s. The histogram of the CCCs is shown in Fig.~\ref{fig: Distribution of CCCs from simulated light curves}, which indicates that the number of segments with CCC $\le$ --0.42 is low (8 out of 700 segments), more segments show lower level of correlation and more than half of the segments show ambiguous correlation.

\begin{table*}
\caption{Percentages of anti-correlations and positive correlations for each spectral state and the whole outburst.}
\begin{center}
\renewcommand{\arraystretch}{1.0}
\label{tab:Percentages of correlation and time lag}
\begin{tabular}{l c c c c c}
\hline \hline
     Correlation/Lag& Hard state& HIMS& SIMS& Soft state& Whole outburst  \\

\noalign{\smallskip}\hline\noalign{\smallskip}
Positive (\%)&16.7&10.0&37.6&14.3&33.6\\
Anti (\%)&0&20.0&6.0&0&6.4\\
Ambiguous (\%)&83.3&70.0&56.4&85.7&60.0\\

\hline \hline

\end{tabular}
\end{center}

\end{table*}

Positive correlations are detected in 47 observations/segments. As shown in Table~\ref{tab:positive correlations}, we can see that positive correlations are detected in every state, viz., hard state, HIMS, SIMS and soft state, but mainly in the SIMS. This is due to the fact that most of the observations in the whole outburst are in the SIMS.
Table~\ref{tab:positive correlations} shows that, in the observations with positive correlations, soft lags are detected in 21 observations/segments located in SIMS, and 24 observations/segments with hard lags are found in hard state, HIMS, SIMS and soft state. The example of positive correlated soft lag and hard lag are shown in the upper and lower panels in Fig.~\ref{fig:positive correlated soft lag and hard lag}, respectively, and the observations with positive correlation do not show spectral pivoting.

6.4\%, 33.6\% and 60.0\% of the total 140 observations/segments display anti-correlation, positive correlation and ambiguous correlation, respectively (see Table~\ref{tab:Percentages of correlation and time lag}).
Fig.~\ref{fig:ambiguous correlation} shows an example for observation with ambiguous correlation, in which the spectral pivoting is not found. It can be seen from Fig.~\ref{fig:lc} that positive correlations and anti-correlations distribute alternately on the light curve, and the number of the observations with correlations begins to decrease from the 40th day of the whole outburst. The distribution of the observations with positive correlation in the HID does not have obvious relation with hardness ratio, while the observations with anti-correlations are distributed in the region with moderate hardness ratio, as shown in Fig.~\ref{fig:HID}.
The time lags in anti-correlation observations are longer than that in positive correlation observations, as shown in Fig.~\ref{fig:time lags vs. MJD}.
To clearly present the evolution of the cross-correlations and time lags along with the spectral states, we show the segment percentage for CCCs and time lags of all the segments for each state in the top panel in Figs.~\ref{fig: percentage of CCCs, distribution of CCCs} and \ref{fig: percentage of time lags, distribution of time lags}, respectively.
The numbers of segments with positive correlations, anti-correlations and ambiguous correlations are shown in the bottom panel of Fig.~\ref{fig: percentage of CCCs, distribution of CCCs}. The numbers of segments with anti-correlated time lag and positive correlated time lag are shown in the bottom panel of Fig.~\ref{fig: percentage of time lags, distribution of time lags}.

\begin{figure*}
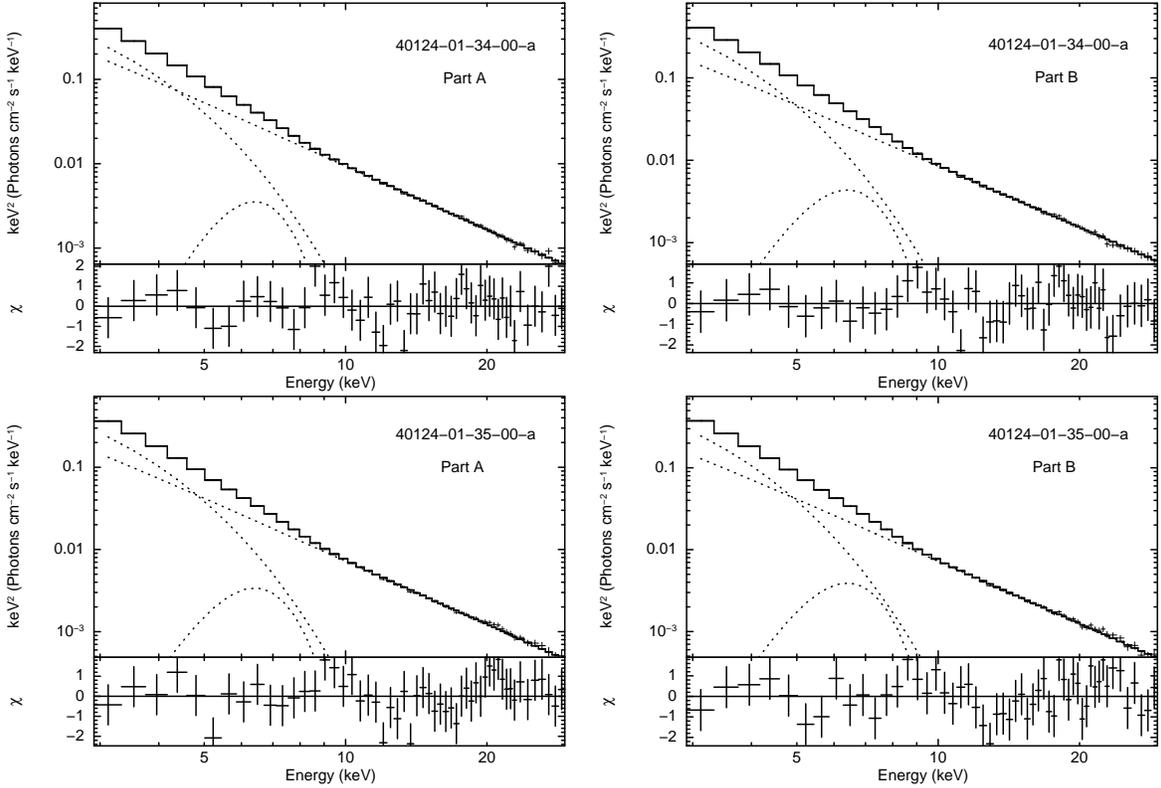

\begin{center}
\includegraphics[width=5.2cm,angle=270,clip]{f28.eps}
\hspace{-0.62em}
\includegraphics[width=5.2cm,angle=270,clip]{f29.eps}
\hspace{-0.62em}
\includegraphics[width=5.2cm,angle=270,clip]{f30.eps}
\hspace{-0.62em}
\includegraphics[width=5.2cm,angle=270,clip]{f31.eps}

\caption{Unfolded spectra and the corresponding $\Delta \chi$ of the two selected segments for spectral analyses, using the model of wabs(diskbb+powerlaw+gauss). A and B correspond to the initial and final parts of the segments.}\label{fig: Unfolded spectra}
\end{center}
\end{figure*}

To study the change of the spectral parameters during the observations with hard lags, we take two segments (40124-01-34-00-a, 40124-01-35-00-a) showing anti-correlated hard lag or positive correlated hard lag to perform the spectra analysis, and we follow \citet{2007ApJ...661.1055S} by extracting the spectrum of two different time intervals (initial and final 300 s) of the observation.
For the PCA spectral fitting, 0.5\% systematic errors are added to the spectra, the hydrogen column density {\it N$_{H}$} is fixed at $0.6 \times 10^{22}$ cm$^{-2}$, the model that consists of a multi-color disk component, a power-law component and a Gaussian component is used, i.e., wabs(diskbb+powerlaw+gauss) \citep{2013ApJ...775...28S}, and the centroid energy of Gaussian component is fixed at 6.4 keV.
For the segment with anti-correlation, the unabsorbed disk flux has an opposite change with the unabsorbed power-law flux.
The ObsID 40124-01-34-00-a shows anti-correlated hard X-ray lag, in which the disk flux increases while the power-law flux decreases, and the spectral analysis in 4U 1630-47 \citep{2007ApJ...661.1055S} shows the same result.
While the segment with positive correlation does not show this phenomenon (see Table~\ref{tab:the unfolded spectra}). From Table~\ref{tab:the unfolded spectra}, we can note that the inner disk radius ($\propto$ $N_{bb}^{1/2}$) has change. Fig.~\ref{fig: Unfolded spectra} shows the unfolded spectra of part A (initial part) and part B (final part) of the two selected segments for spectral analyses.

\section[]{DISCUSSION}

Positive correlations are common in XRBs and easily understood. Hard X-ray emission comes from the inverse Compton scattering (ICS) of soft seed photons, so the hard X-ray flux changes along with the increase or decrease of soft X-ray flux in general. Therefore, the following discussion will concentrate on anti-correlations.

\subsection{Comparison with other XRBs}

For XTE J1859+226, anti-correlations are detected in the IMS, viz., HIMS and SIMS, and the most are in the SIMS. Anti-correlations are also detected in the IMS of other five BHXBs which are GRS 1915+105, XTE J1550-564, 4U 1630-47, H1743-322 and GX 339-4. However, for BHXB Cyg X-3, anti-correlations are only detected in the hard  state. In a word, anti-correlations are not found in the soft state of BHXBs.

However, anti-correlations are detected in every branch or state of NS LMXBs.
Z-sources show Z-shaped tracks in CCDs along three branches, which are called horizontal branch (HB), normal branch (NB), and flaring branch (FB), respectively.
The atoll sources show island state (IS) and banana state (BS) in CCDs, and BS can be subdivided into three states, viz., lower left banana (LLB), lower banana (LB), and upper banana (UB) states.
Anti-correlations are found in HB, NB and FB of three Z-sources, viz., Cyg X-2 \citep{2008ApJ...677..461L}, GX 5-1 \citep{2012ApJS..200...16S} and XTE J1701-462 \citep{2014MNRAS.440.3726W}.
For the atoll source, the anti-correlated observations are detected in the LB and UB of 4U 1735-44 \citep{2013AJ....146...60L}, and in the IS, LLB and UB of 4U 1608-52 \citep{2014AJ....147...67L}.
In summary, anti-correlations are not detected in the soft state of BHXBs, but in every branch or state of NS LMXBs. This may be due to the spectra being dominated by the soft component in the soft state of BHXBs, while the soft seed photons in BHXBs only originate from the accretion disk but from both the accretion disk and the surface of the NS for NS LMXBs.

For the observations with anti-correlations in BHXBs, hard lags are generally found.
For XTE J1859+226, the anti-correlated hard lags are detected in the SIMS, and that lags range from $\sim$ 204 to $\sim$ 953 seconds.
For Cyg X-3, the anti-correlated hard lags are detected in the hard state, which range from $\sim$ 620 to $\sim$ 950 seconds \citep{2004ApJ...616L.143C}.
The anti-correlated hard lags are detected in the IMS of GRS 1915+105 \citep{2005ApJ...631.1072C}, XTE 1550-564 \citep{2007ApJ...661.1055S} and H1743-322 \citep{2009RAA.....9..901S}, which range from $\sim$ 128 to $\sim$ 1600 seconds, $\sim$ 132 to $\sim$ 376 seconds and $\sim$ 442 to $\sim$ 1591 seconds, respectively.
From the above results, it is easy to note that the ranges of anti-correlated hard lags detected in different BHXBs are similar.

Anti-correlated soft lags are first found in the IMS of GX 339-4 which range from $\sim$ 95 to $\sim$ 1068 seconds \citep{2010ApJ...725.1317S}.
XTE J1859+226 is the second BHXB that shows anti-correlated soft lags that range from $\sim$ 120 to $\sim$ 718 seconds.
The ranges of anti-correlated soft lags detected in these two BHXBs do not differ significantly.
Table~\ref{tab:ranges of time lags} shows all the ranges for time lags detected in BHXBs.

\begin{figure*}
\centering

\includegraphics[width=16cm]{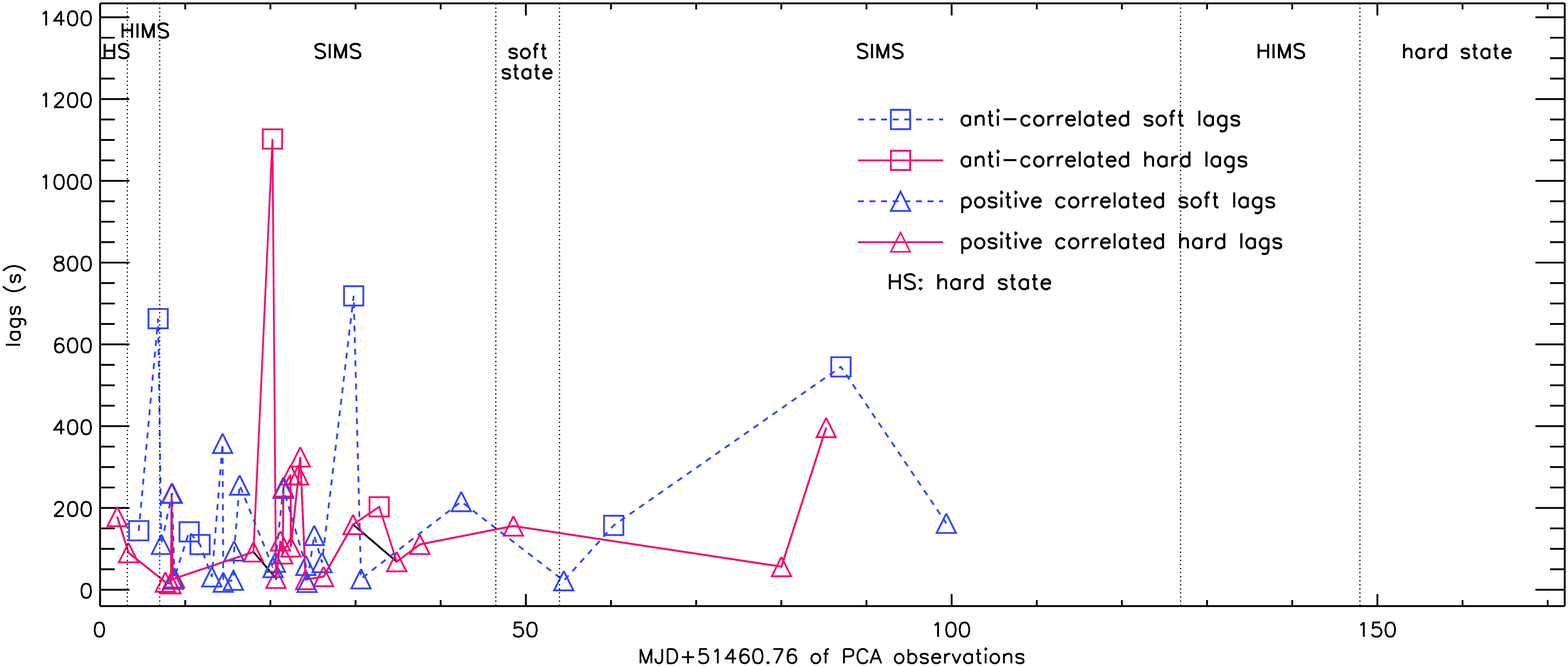}

\caption{Soft lags and hard lags vs. MJD for the anti-correlated soft lag, anti-correlated hard lag, positive correlated soft lag and positive correlated hard lag, as shown in the legend.}\label{fig:time lags vs. MJD}

\end{figure*}

\begin{table*}
\footnotesize
\caption{XTE J1859+226 and other twelve XRBs in which long time-scale lags are detected and their ranges of time lags.}
\begin{center}
\renewcommand{\arraystretch}{1.0}
\label{tab:ranges of time lags}

\begin{tabular}{l c c c c}
\hline \hline
     Object&Range of AS Lag (s)&Range of AH Lag (s)&Range of PS Lag (s)& Range of PH Lag (s)   \\

\hline

\multicolumn{5}{c}{Z-source}\\

\hline

Cyg X-2 (1)&-35$\pm$25 $\sim$ -184$\pm$29&46$\pm$14 $\sim$ 376$\pm$25&...&...\\
GX 5-1 (2)&-160$\pm$70 $\sim$ -920$\pm$102&31$\pm$12 $\sim$ 748$\pm$62&...&...\\
XTE J1701-462 (3)&-29$\pm$15 $\sim$ -1278$\pm$30&54$\pm$14 $\sim$ 1138$\pm$26&-9$\pm$6 $\sim$ -1334$\pm$15&17$\pm$10 $\sim$ 1300$\pm$23\\
GX 349+2 (4)&-66$\pm$10 $\sim$ -6107$\pm$46&85$\pm$6 $\sim$ 1177$\pm$57&-3.9$\pm$3.2 $\sim$ -286$\pm$40&2.8$\pm$1.8 $\sim$ 136$\pm$27\\

\hline

\multicolumn{5}{c}{atoll source}\\

\hline

4U 1735-44 (5)&-26$\pm$25 $\sim$ -350$\pm$10&63$\pm$33 $\sim$ 152$\pm$28&-14$\pm$5 $\sim$ -337$\pm$19&25$\pm$15 $\sim$ 368$\pm$24\\
4U 1608-52 (6)&-100$\pm$102&168$\pm$32 $\sim$ 628$\pm$43&-13$\pm$10 $\sim$ -50$\pm$44&19$\pm$19 $\sim$ 248$\pm$77\\

\hline

\multicolumn{5}{c}{BHXB}\\

\hline

Cyg X-3 (7)&...&620$\pm$70 $\sim$ 950$\pm$60&...&...\\
GRS 1915+105 (8)&...&128$\pm$30 $\sim$ 1600$\pm$200&...&...\\
XTE 1550-564 (9)&...&132$\pm$9 $\sim$ 376$\pm$14&...&...\\
4U 1630-47 (9)&...&363$\pm$48&...&...\\
H1743-322 (10)&...&442$\pm$38 $\sim$ 1591$\pm$58&...&...\\
GX 339-4 (11)&-95$\pm$40 $\sim$ -1068$\pm$62& ...& ...&...\\
\bfseries XTE J1859+226&\bfseries -120$\pm$16 $\sim$ -718$\pm$75&\bfseries 204$\pm$87 $\sim$ 953$\pm$65&\bfseries -12.6$\pm$3.7 $\sim$ -362$\pm$41&\bfseries 12.1$\pm$3.7 $\sim$ 403$\pm$62\\
\hline \hline

\end{tabular}
\end{center}
\footnotesize{
\begin{quote}{\bf Notes.}
AS Lag: anti-correlated soft lag

AH Lag: anti-correlated hard lag

PS Lag: positive correlated soft lag

PH Lag: positive correlated hard lag

References: (1) \citet{2008ApJ...677..461L}, (2) \citet{2012ApJS..200...16S}, (3) \citet{2014MNRAS.440.3726W}, (4) \citet{2016MNRAS.455.2959D}, (5) \citet{2013AJ....146...60L}, (6) \citet{2014AJ....147...67L}, (7) \citet{2004ApJ...616L.143C}, (8) \citet{2005ApJ...631.1072C}, (9) \citet{2007ApJ...661.1055S}, (10) \citet{2009RAA.....9..901S} and (11) \citet{2010ApJ...725.1317S}.
\end{quote}
}
\end{table*}

We notice that the time lags in anti-correlation observations of XTE J1859+226 are longer than that in positive correlation observations (see Fig.~\ref{fig:time lags vs. MJD}). This phenomenon is also detected in atoll source 4U 1608-52 \citep{2014AJ....147...67L} and Z-source GX 349+2 \citep{2016MNRAS.455.2959D}.
However, for atoll source 4U 1735-44 \citep{2013AJ....146...60L} and Z-source XTE J1701-462 \citep{2014MNRAS.440.3726W}, the length of time lags in positive correlation observations does not show obvious difference from that in anti-correlation observations.

It is interesting to note that the proportion of anti-correlated soft lags and anti-correlated hard lags does not show obvious relation with the category of source. Of the observations of XTE J1859+226 that show anti-correlations, the percentage of observations with soft lags (78\%) is more than that with hard lags (22\%), as shown in the bottom panel of Fig.~\ref{fig: percentage of time lags, distribution of time lags}. GX 349+2 (Z-source) shows the same phenomenon \citep{2016MNRAS.455.2959D}, while the opposite is detected in Z-source Cyg X-2 \citep{2008ApJ...677..461L}, Z-source GX 5-1 \citep{2012ApJS..200...16S}, Z-source XTE J1701-462 \citep{2014MNRAS.440.3726W}, atoll source 4U 1735-44 \citep{2013AJ....146...60L} and atoll source 4U 1608-52 \citep{2014AJ....147...67L}.

\subsection{Explanation of the Results in XTE J1859+226}

Model-independent spectral pivoting is detected in the anti-correlated observations of BHXB XTE J1859+226, this spectral pivoting is the strongest evidence for the anti-correlation \citep{2003ApJ...593..452C, 2004ApJ...616L.143C}, and this kind of spectral pivoting has been detected in the anti-correlated observations of six BHXBs (Cyg X-3 \citep{2004ApJ...616L.143C}, GRS 1915+105 \citep{2005ApJ...631.1072C}, XTE J1550-564 and 4U 1630-47 \citep{2007ApJ...661.1055S}, H1743-322 \citep{2009RAA.....9..901S} and GX 339-4 \citet{2010ApJ...725.1317S}) and five NS LMXBs (Cyg X-2 \citep{2008ApJ...677..461L}, GX 5-1 \citep{2012ApJS..200...16S}, 4U 1735-44 \citep{2013AJ....146...60L}, 4U 1608-52 \citep{2014AJ....147...67L} and XTE J1701-462 \citep{2014MNRAS.440.3726W}). However, spectral pivoting is not detected in the positively correlated observations of XTE J1859+226, and the same result is found in the above five NS LMXBs. This kind of model-independent spectral pivoting behavior might imply that the radiation mechanism and spectral properties are different in the anti-correlated observations and positively correlated observations.

\begin{figure}
\begin{center}
\includegraphics[width=8cm, height=6cm]{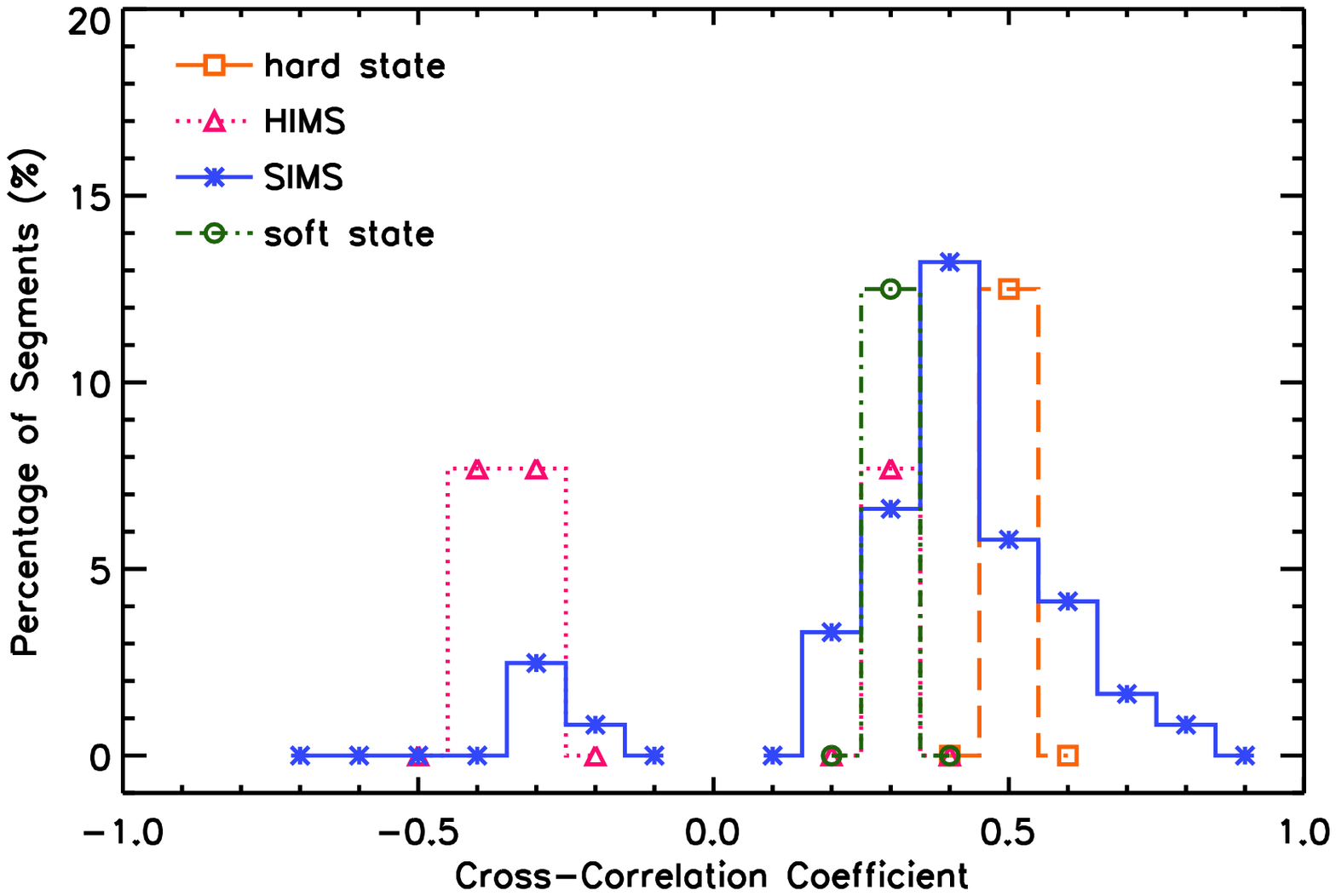}
\includegraphics[width=8cm, height=6cm]{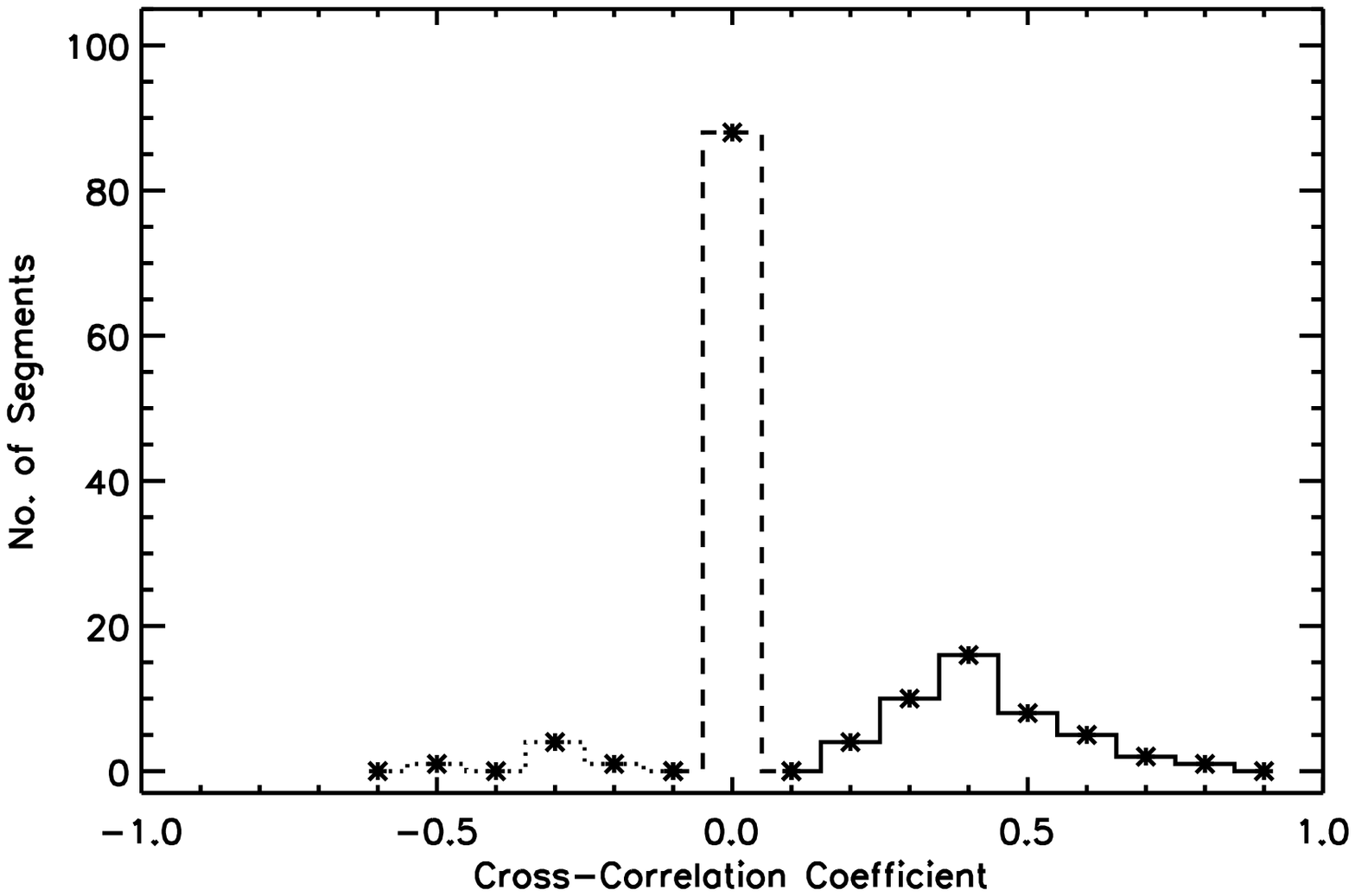}
\caption{Top panel: the percentage of segments with CCCs. Bottom panel: the CCCs of the anti-correlations (dotted line), positive correlations (solid line) and ambiguous correlations (dashed line) vs. the number of segments.}\label{fig: percentage of CCCs, distribution of CCCs}
\end{center}
\end{figure}

\begin{figure}
\begin{center}
\includegraphics[width=8cm, height=6cm]{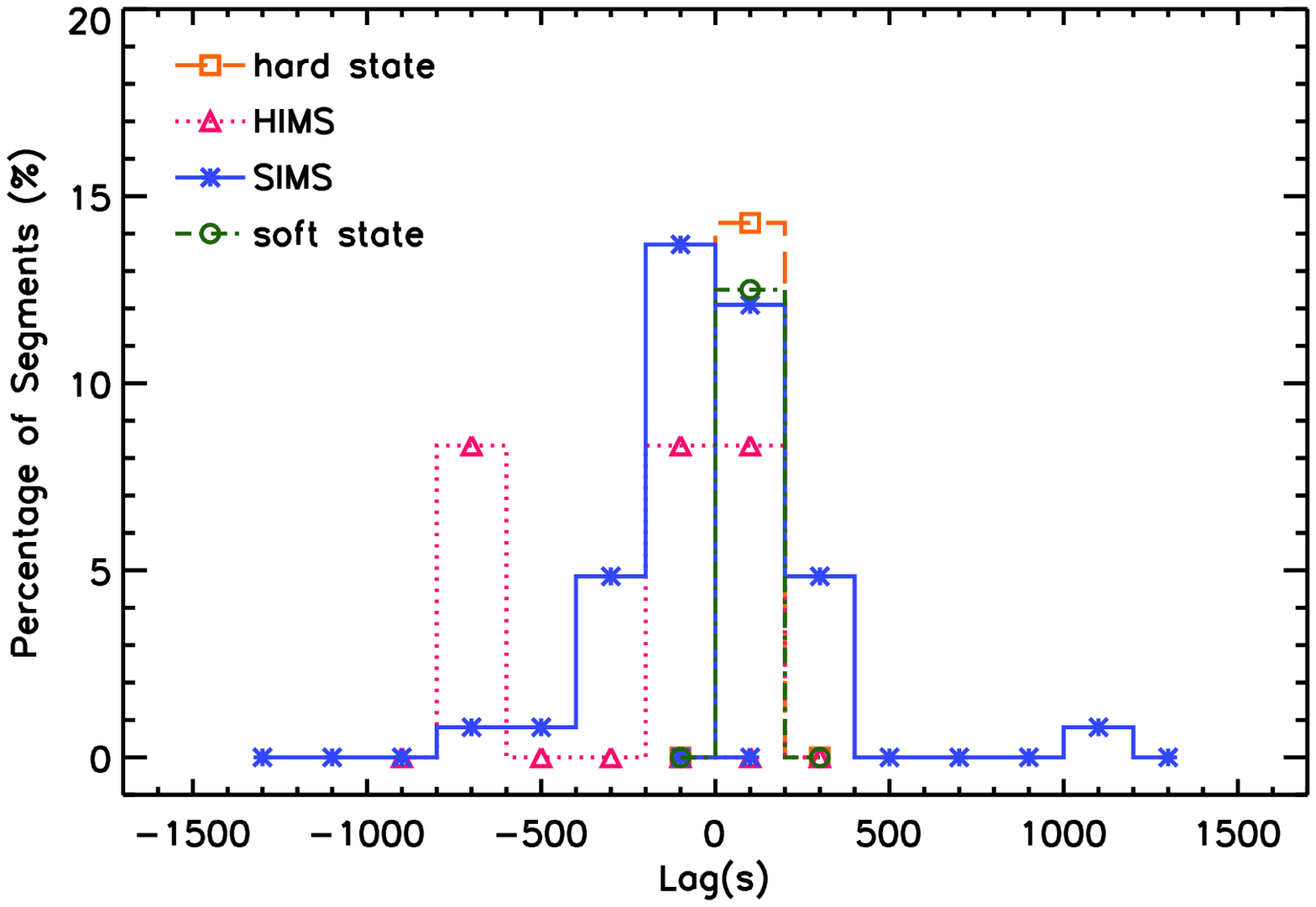}
\includegraphics[width=8cm, height=6cm]{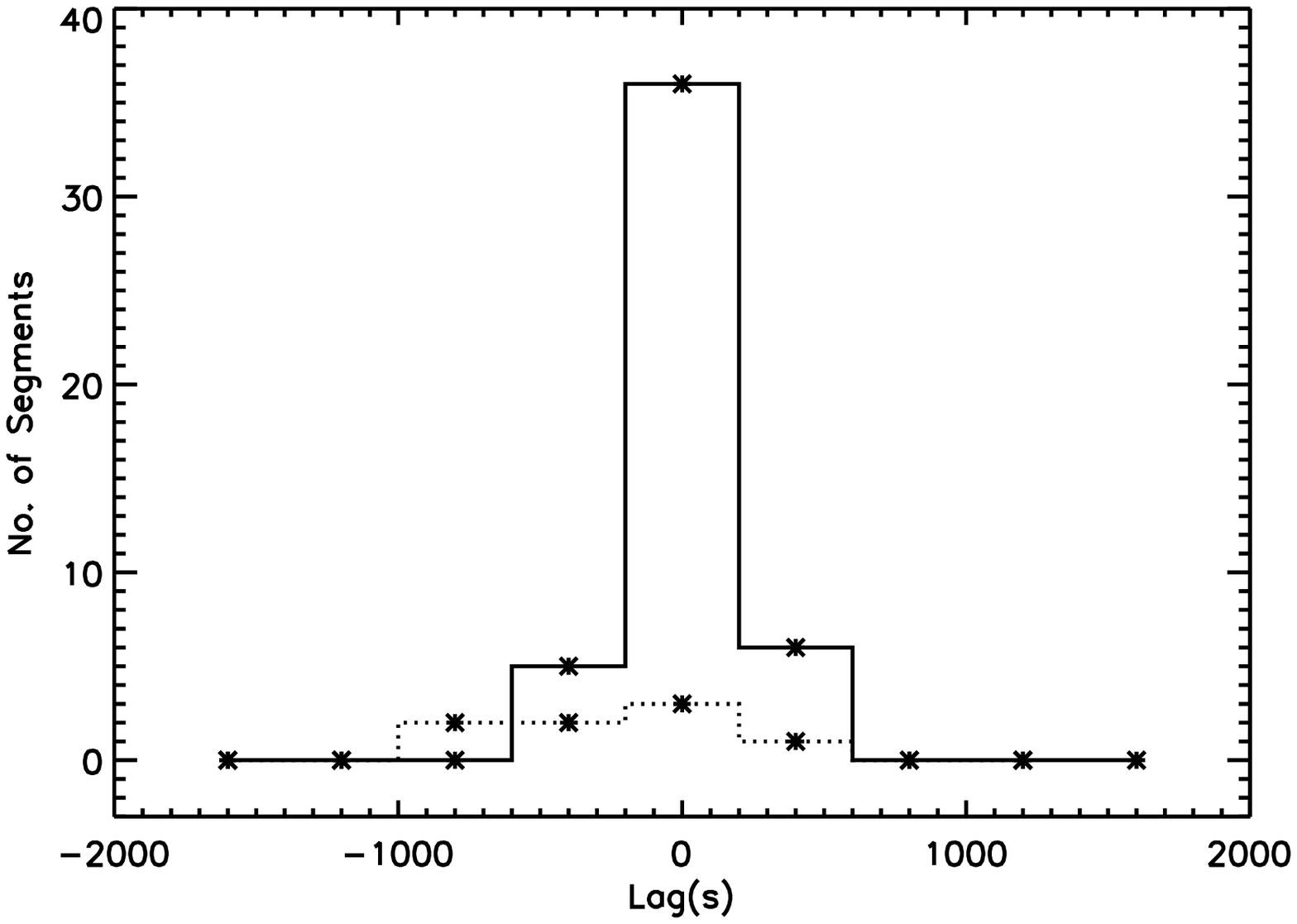}

\caption{Top panel: percentage of segments vs. time lags. Bottom panel: the time lags of the anti-correlations (dotted line), positive correlations (solid line) vs. the number of segments.}\label{fig: percentage of time lags, distribution of time lags}

\end{center}
\end{figure}

For XTE J1859+226, both anti-correlations and positive correlations are found in the first 100 days (out of 166 days) when the luminosity is high. The number of observations with anti-correlations or positive correlations begins to decrease from the 40th day of the whole outburst (see Fig.~\ref{fig:lc}).
These results imply that the distribution of cross-correlations may be related to the high luminosity of the BHXB. Positive correlations do not show clear evolution with hardness ratio in the HID, whereas anti-correlations are detected in the observations with moderate hardness ratio, as illustrated by the Fig.~\ref{fig:HID}. Therefore, anti-correlations may be related to the moderate hardness ratio in BHXB.

Anti-correlated lags detected in XTE J1859+226, which range from $\sim$ 120 to $\sim$ 953 seconds, are long time-scale lags (longer than 1 second).
The short time-scale X-ray lags, of less than 1 second, observed in XRBs can be explained by Comptonization model \citep{2001ApJ...549L.229L} and shot model \citep{1987ApJ...319L..13V, 1989Natur.342..773M, 1999ApJ...517..355N}.
However, these models can not explain the long-term time lags detected in BHXB, which range from hundreds to thousands of seconds.
The truncated accretion disk model is invoked to interpret the anti-correlated long time-scale hard X-ray lags detected in the hard state of BHXB Cyg X-3 \citep{2004ApJ...616L.143C}.
In this model, the accretion disk is truncated far away from the compact object. The high-temperature Comptonizing cloud is constrained to the area between the truncated disk and the compact object. The accretion disk front moves inward (outward), leading to a decrease (increase) of the Comptonizing cloud.
Any changes in the accretion disk will result in opposite changes in the Comptonizing cloud resulting in anti-correlated hard X-ray lags corresponding to the viscous timescale of the readjustment of the accretion disk.

The truncated accretion disk model is widely used to explain similar hard X-ray lags in BHXBs (GRS 1915+105 \citep{2005ApJ...631.1072C}, XTE J1550-564 and 4U 1630-47 \citep{2007ApJ...661.1055S}, H1743-322 \citep{2009RAA.....9..901S}) and NS LMXBs (Cyg X-2 (Z-source) \citep{2008ApJ...677..461L}, GX 5-1 (Z-source) \citep{2012ApJS..200...16S}, 4U 1735-44 (atoll source) \citep{2013AJ....146...60L}, 4U 1608-52 (atoll source) \citep{2014AJ....147...67L}, XTE J1701-462 (Z-source) \citep{2014MNRAS.440.3726W}). Spectral analysis shows that the inner disk radius ($\propto$ $N_{bb}^{1/2}$) is changing, and this could be the evidence of the movement of accretion disk front (see Table~\ref{tab:the unfolded spectra}). Therefore, we suggest that the anti-correlated hard X-ray lags detected in XTE J1859+226 are also produced by that mechanism.

{\begin{table*}

\caption{Best-fitting parameters for part A (initial part) and part B (final part) spectral of the two selected segments for spectral analyses.}
\vspace{2.mm}

\renewcommand{\arraystretch}{1.5} 

\begin{center}
\renewcommand{\footnoterule}{}
\label{tab:the unfolded spectra}
\begin{tabular}{ccccc}
\hline
Parameters&\multicolumn{2}{c}{40124-01-34-00-a} &\multicolumn{2}{c}{40124-01-35-00-a}\\
\hline
&A&B&A&B\\

\hline
$kT_{in}$$^{a}$&$0.84_{-0.01}^{+0.01}$ &$0.85_{-0.01}^{+0.01}$ &$0.83_{-0.01}^{+0.01}$ &$0.85_{-0.01}^{+0.01}$\\
$N_{bb}$$^{b}$&$752_{-38}^{+39}$ &$840_{-42}^{+45}$ &$802_{-40}^{+43}$ &$896_{-44}^{+46}$\\
$\Gamma_{Pl}$$^{c}$&$2.52_{-0.02}^{+0.03}$ &$2.48_{-0.01}^{+0.01}$ &$2.56_{-0.02}^{+0.02}$ &$2.47_{-0.02}^{+0.02}$\\
Disk flux$^{d}$&$1.88_{-0.05}^{+0.05}$ &$2.14_{-0.07}^{+0.08}$ &$2.12_{-0.03}^{+0.04}$ &$2.18_{-0.05}^{+0.04}$\\
Power-law flux &$4.28_{-0.06}^{+0.06}$ &$3.85_{-0.08}^{+0.08}$ &$3.11_{-0.05}^{+0.06}$ &$3.11_{-0.06}^{+0.06}$\\
$Flux_{total}$ &6.25 & 6.09 &5.33 & 5.39\\
Disk flux/$Flux_{total}$$^{e}$ &30.08\% & 35.14\% &39.77\% & 40.45\% \\
$\chi^{2}$/dof&42/51 & 52/51 &53/51 & 52/51\\

\hline
\hline

\end{tabular}
\end{center}
\footnotesize{
\begin{quote}{\bf Notes.}
The model wabs(diskbb+powerlaw+gauss) is used, and all the quoted errors are at a 90 percent confidence level.

\leftline{$^{a}$Inner disk temperature of diskbb model.}
\leftline{$^{b}$Normalization of discbb model.}
\leftline{$^{c}$Power-law index.}
\leftline{$^{d}$For all the models, the flux unit is $10^{-9}$ergs $cm^{-2} s^{-1}$, and fluxes are unabsorbed in 2.8$-$30 keV energy band.}
\leftline{$^{e}$The ratio of the disk flux to the total flux.}
\end{quote}
}

\end{table*}}

Anti-correlated soft X-ray lags can also be interpreted in the framework of the truncated accretion disk model.
\citet{2007ApJ...666..368L} suggested that the fluctuation produced in the innermost region of the BH accretion disk expands and propagates to the outer region of accretion disk, that fluctuation heats up the inner material and disturbs the outer material, and that fluctuation lasts within thousands of seconds for the accretion disk with Stellar-Mass Black Hole.
\citet{2008ApJ...677..461L} invoked that fluctuation to explain the anti-correlated soft lags detected in Cyg X-2 (Z-source), and that fluctuation mechanism was also adopted to interpret similar soft lags detected in Z-source GX 5-1 \citep{2012ApJS..200...16S}, atoll source 4U 1735-44 \citep{2013AJ....146...60L} and Z-source XTE J1701-462 \citep{2014MNRAS.440.3726W}.
\citet{2010ApJ...725.1317S} detected anti-correlated long time-scale soft X-ray lags in the IMS of BHXB GX 339-4 and they suggested that the temporal and spectral variations during the soft lag timescale could be explained with the framework of the truncated accretion disk model.
XTE J1859+226 is the second BHXB in which the anti-correlated long time-scale soft X-ray lags are detected.
We suggest that, in the framework of the truncated accretion disk model, this fluctuation mechanism can explain the anti-correlated long time-scale soft X-ray lags detected in BHXB as well.
Considering the anti-correlated long time-scale X-ray lags are detected in IMS of XTE J1859+226 but never in the soft state of BHXB, we suggest that the accretion disk is truncated at the IMS of BHXB but may be not truncated at soft state.

The proportion of multi-color disk component in the segment with positive correlation is higher than that in the segment with anti-correlation (see Table~\ref{tab:the unfolded spectra}). Therefore, soft seed thermal photons emitted in the accretion disk dominate the cross-correlations between the soft and hard X-rays during the observations with positive correlation, i.e., the decrease or increase of soft seed thermal photons will lead to a same change in the hard X-ray emission.
The relative changes in inner disk radius observed from the spectral fits are just close to 90\% confidence level (Table~\ref{tab:the unfolded spectra}) and resolved spectral observations are required in order to constrain the changes during the anti-correlated observations.
In addition, the change of inner disk radius ($\propto$ $N_{bb}^{1/2}$) has the similar amplitude during the segments with positive correlation and anti-correlation (see Table~\ref{tab:the unfolded spectra}).
We suggest that the positive correlated hard lags may correspond to the viscous timescale of the readjustment of the accretion disk as anti-correlated hard lags, and the positive correlated soft lags may result from the fluctuation as anti-correlated soft lags for BHXBs.

\section[]{CONCLUSION}

Using the data from RXTE for the BHXB XTE J1859+226 during its entire 1999$-$2000 outburst, we perform a systematic study of the CCF and time lag between the soft and hard X-rays.
For the first time, positive correlated soft lags and hard lags detected in BHXB are reported. Anti-correlated soft lags are second observed in BHXB, and anti-correlated hard lags are also observed in BHXB XTE J1859+226.
Anti-correlations are detected in every branch or state of NS LMXBs, but never detected in the soft state of BHXBs. This may be due to the soft seed photons in NS LMXBs and BHXBs being emitted from different regions.
The timescales of anti-correlated hard lags detected in different BHXBs do not show obvious difference, where the timescales range from several hundreds to more than one thousand seconds.
The ranges of anti-correlated soft lags detected in XTE J1859+226 and BHXB GX 339-4 do not have obvious difference.
In XTE J1859+226, the time lags detected in anti-correlation observations are longer than that in positive correlation observations.
In the observations with anti-correlations, most of the lags are soft lags.
The proportion of the observations with anti-correlated soft lags and hard lags does not display obvious relation with the source type.
Anti-correlations and positive correlations and their respective soft and hard X-ray lags are only detected in the first 100 days of the outburst when the luminosity or the accretion rate are high implying that the cross-correlations between the soft and hard X-rays may be related to the high luminosity.
We suggest that anti-correlated long time-scale X-ray hard lag and soft lag can be explained by combination of the truncated accretion disk model and a fluctuation that propagates from the inner disk to the outer disk.
The anti-correlated long time-scale X-ray lags are never found in the soft state of BHXBs probably because the accretion disk is not truncated at soft state.

\section*{Acknowledgements}

We thank the anonymous referee for her or his constructive comments and suggestions, which helped us to carry out this research deeply and improve the presentation of this paper. This research has made use of the data obtained through the High Energy Astrophysics Science Archive Research Center (HEASARC) On-line Service, provided by NASA/Goddard Space Flight Center (GSFC). This work is partially supported by the National Program on Key Research and Development Project (Grant Nos. 2016YFA0400804 and 2016YFA0400803) and the National Key Basic Research Program of China (973 Program 2015CB857100). This work is also partially supported by the Natural Science Foundation of China (Grant Nos. 11303047, 11573059 and 11673023) and the 2014 Project of Xinjiang Uygur Autonomous Region of China for Flexibly Fetching in Upscale Talents.

\label{lastpage}


\begin{thebibliography}{}

\bibitem[\protect\citeauthoryear{Belloni}{2010}]{2010LNP...794...53B} Belloni T.~M., 2010, LNP, 794, 53

\bibitem[\protect\citeauthoryear{Choudhury \& Rao}{2004}]{2004ApJ...616L.143C} Choudhury M., Rao A.~R., 2004, ApJ, 616, L143

\bibitem[\protect\citeauthoryear{Choudhury et al.}{2005}]{2005ApJ...631.1072C} Choudhury M., Rao A.~R., Dasgupta S., Pendharkar J., Sriram K., Agrawal V.~K., 2005, ApJ, 631, 1072

\bibitem[\protect\citeauthoryear{Choudhury et al.}{2003}]{2003ApJ...593..452C} Choudhury M., Rao A.~R., Vadawale S.~V., Jain A.~K., 2003, ApJ, 593, 452

\bibitem[\protect\citeauthoryear{Colpi et al.}{2009}]{2009ASSL..359.....C} Colpi M., Casella P., Gorini V., Moschella U., Possenti A., 2009, ASSL, 359,

\bibitem[\protect\citeauthoryear{Corral-Santana et al.}{2011}]{2011MNRAS.413L..15C} Corral-Santana J.~M., Casares J., Shahbaz T., Zurita C., Mart{\'{\i}}nez-Pais I.~G., Rodr{\'{\i}}guez-Gil P., 2011, MNRAS, 413, L15

\bibitem[\protect\citeauthoryear{Cui et al.}{2000}]{2000ApJ...535L.123C} Cui W., Shrader C.~R., Haswell C.~A., Hynes R.~I., 2000, ApJ, 535, L123

\bibitem[\protect\citeauthoryear{Debnath et al.}{2008}]{2008BASI...36..151D} Debnath D., Chakrabarti S.~K., Nandi A., Mandal S., 2008, BASI, 36, 151

\bibitem[\protect\citeauthoryear{Ding et al.}{2016}]{2016MNRAS.455.2959D} Ding G.~Q., Zhang W.~Y., Wang Y.~N., Li Z.~B., Qu J.~L., Huang C.~P., 2016, MNRAS, 455, 2959

\bibitem[\protect\citeauthoryear{Done, Gierli{\'n}ski, \& Kubota}{2007}]{2007A&ARv..15....1D} Done C., Gierli{\'n}ski M., Kubota A., 2007, A\&ARv, 15, 1

\bibitem[\protect\citeauthoryear{Dunn et al.}{2010}]{2010MNRAS.403...61D} Dunn R.~J.~H., Fender R.~P., K{\"o}rding E.~G., Belloni T., Cabanac C., 2010, MNRAS, 403, 61

\bibitem[\protect\citeauthoryear{Farinelli et al.}{2013}]{2013MNRAS.428.3295F} Farinelli R., et al., 2013, MNRAS, 428, 3295

\bibitem[\protect\citeauthoryear{Hasinger \& van der Klis}{1989}]{1989A&A...225...79H} Hasinger G., van der Klis M., 1989, A\&A, 225, 79

\bibitem[\protect\citeauthoryear{Homan \& Belloni}{2005}]{2005Ap&SS.300..107H} Homan J., Belloni T., 2005, Ap\&SS, 300, 107

\bibitem[\protect\citeauthoryear{Hynes et al.}{2002}]{2002MNRAS.331..169H} Hynes R.~I., Haswell C.~A., Chaty S., Shrader C.~R., Cui W., 2002, MNRAS, 331, 169

\bibitem[\protect\citeauthoryear{Lee, Misra, \& Taam}{2001}]{2001ApJ...549L.229L} Lee H.~C., Misra R., Taam R.~E., 2001, ApJ, 549, L229

\bibitem[\protect\citeauthoryear{Lei et al.}{2008}]{2008ApJ...677..461L} Lei Y.~J., et al., 2008, ApJ, 677, 461-472

\bibitem[\protect\citeauthoryear{Lei et al.}{2013}]{2013AJ....146...60L} Lei Y.-J., et al., 2013, AJ, 146, 60

\bibitem[\protect\citeauthoryear{Lei et al.}{2014}]{2014AJ....147...67L} Lei Y.-J., et al., 2014, AJ, 147, 67

\bibitem[\protect\citeauthoryear{Lewin \& van der Klis}{2006}]{2006csxs.book.....L} Lewin W.~H.~G., van der Klis M., 2006, csxs.book, 39,

\bibitem[\protect\citeauthoryear{Li, Xue, \& Lu}{2007}]{2007ApJ...666..368L} Li S.-L., Xue L., Lu J.-F., 2007, ApJ, 666, 368

\bibitem[\protect\citeauthoryear{Markoff \& Nowak}{2004}]{2004ApJ...609..972M} Markoff S., Nowak M.~A., 2004, ApJ, 609, 972

\bibitem[\protect\citeauthoryear{Markoff, Nowak, \& Wilms}{2005}]{2005ApJ...635.1203M} Markoff S., Nowak M.~A., Wilms J., 2005, ApJ, 635, 1203

\bibitem[\protect\citeauthoryear{Markwardt, Marshall, \& Swank}{1999}]{1999IAUC.7274....2M} Markwardt C.~B., Marshall F.~E., Swank J.~H., 1999, IAUC, 7274, 2

\bibitem[\protect\citeauthoryear{McClintock \& Remillard}{2006}]{2006csxs.book..157M} McClintock J.~E., Remillard R.~A., 2006, csxs.book, 39, 157

\bibitem[\protect\citeauthoryear{Miyamoto \& Kitamoto}{1989}]{1989Natur.342..773M} Miyamoto S., Kitamoto S., 1989, Natur, 342, 773

\bibitem[\protect\citeauthoryear{Motta, Belloni, \& Homan}{2009}]{2009MNRAS.400.1603M} Motta S., Belloni T., Homan J., 2009, MNRAS, 400, 1603

\bibitem[\protect\citeauthoryear{Nandi et al.}{2012}]{2012A&A...542A..56N} Nandi A., Debnath D., Mandal S., Chakrabarti S.~K., 2012, A\&A, 542, A56

\bibitem[\protect\citeauthoryear{Nowak, Wilms, \& Dove}{1999}]{1999ApJ...517..355N} Nowak M.~A., Wilms J., Dove J.~B., 1999, ApJ, 517, 355

\bibitem[\protect\citeauthoryear{Radhika \& Nandi}{2014}]{2014AdSpR..54.1678R} Radhika D., Nandi A., 2014, AdSpR, 54, 1678

\bibitem[\protect\citeauthoryear{Rau \& Greiner}{2003}]{2003A&A...397..711R} Rau A., Greiner J., 2003, A\&A, 397, 711

\bibitem[\protect\citeauthoryear{Remillard}{2005}]{2005AIPC..797..231R} Remillard R.~A., 2005, AIPC, 797, 231

\bibitem[\protect\citeauthoryear{Remillard \& McClintock}{2006}]{2006ARA&A..44...49R} Remillard R.~A., McClintock J.~E., 2006, ARA\&A, 44, 49

\bibitem[\protect\citeauthoryear{Shaposhnikov \& Titarchuk}{2009}]{2009ApJ...699..453S} Shaposhnikov N., Titarchuk L., 2009, ApJ, 699, 453

\bibitem[\protect\citeauthoryear{Sriram et al.}{2007}]{2007ApJ...661.1055S} Sriram K., Agrawal V.~K., Pendharkar J.~K., Rao A.~R., 2007, ApJ, 661, 1055

\bibitem[\protect\citeauthoryear{Sriram, Agrawal, \& Raghurama Rao}{2009}]{2009RAA.....9..901S} Sriram K., Agrawal V.~K., Raghurama Rao A., 2009, RAA, 9, 901

\bibitem[\protect\citeauthoryear{Sriram, Choi, \& Rao}{2012}]{2012ApJS..200...16S} Sriram K., Choi C.~S., Rao A.~R., 2012, ApJS, 200, 16

\bibitem[\protect\citeauthoryear{Sriram, Rao, \& Choi}{2010}]{2010ApJ...725.1317S} Sriram K., Rao A.~R., Choi C.~S., 2010, ApJ, 725, 1317

\bibitem[\protect\citeauthoryear{Sriram, Rao, \& Choi}{2013}]{2013ApJ...775...28S} Sriram K., Rao A.~R., Choi C.~S., 2013, ApJ, 775, 28

\bibitem[\protect\citeauthoryear{Timmer \& Koenig}{1995}]{1995A&A...300..707T} Timmer J., Koenig M., 1995, A\&A, 300, 707

\bibitem[\protect\citeauthoryear{van der Klis et al.}{1987}]{1987ApJ...319L..13V} van der Klis M., Hasinger G., Stella L., Langmeier A., van Paradijs J., Lewin W.~H.~G., 1987, ApJ, 319, L13

\bibitem[\protect\citeauthoryear{Wang et al.}{2014}]{2014MNRAS.440.3726W} Wang Y.~N., Lei Y.~J., Ding G.~Q., Qu J.~L., Ge M.~Y., Zhang C.~M., Chen L., Ma X., 2014, MNRAS, 440, 3726

\bibitem[\protect\citeauthoryear{Wood et al.}{1999}]{1999IAUC.7274....1W} Wood A., Smith D.~A., Marshall F.~E., Swank J., 1999, IAUC, 7274, 1

\bibitem[\protect\citeauthoryear{Zurita et al.}{2002}]{2002MNRAS.334..999Z} Zurita C., et al., 2002, MNRAS, 334, 999

\end{thebibliography}
\end{document}